\documentclass[useAMS,usenatbib]{mn2e}

\usepackage{times}

\usepackage{graphicx}

\usepackage{multirow}

\usepackage{natbib}

\newcommand{\avg}[1]{\ensuremath{\langle #1 \rangle}}
\newcommand{\Ang}{\mbox{ \AA}}

\begin{document}
	
\title[What does $f_{Ly\alpha}$ tell us about reionization?]{What do observations of the Lyman $\alpha$ fraction tell us about reionization?}

\author[J. Taylor and A. Lidz]{Jessie Taylor$^1$\thanks{tfj@sas.upenn.edu} \& Adam Lidz$^{1}$\\
${^1}$ Department of Physics and Astronomy, University of Pennsylvania, 209 South 33rd Street, Philadelphia, PA 19104, USA\\}
	
\date{\today}
	
\maketitle

\begin{abstract}
An appealing approach for studying the reionization history of the
Universe is to measure the redshift evolution of the Lyman $\alpha$ (Ly$\alpha$)
fraction, the percentage of Lyman-break selected galaxies that emit
appreciably in the Ly$\alpha$ line. This fraction is expected to
fall-off towards high redshift as the intergalactic medium becomes significantly neutral, and the
galaxies' Ly$\alpha$ emission is progressively attenuated.
Intriguingly, early
measurements with this technique suggest a strong drop in the
Ly$\alpha$ fraction near $z \sim 7$. Previous work concluded that this requires a surprisingly neutral intergalactic medium --
with neutral hydrogen filling more than 50 per cent of the volume of the
Universe -- at this redshift. We model the  evolving Ly$\alpha$
fraction using cosmological simulations of the reionization
process. Before reionization completes, the simulated Ly$\alpha$ fraction has
large spatial fluctuations owing to the inhomogeneity of 
reionization. Since existing measurements of the Ly$\alpha$ fraction
span relatively small regions on the sky, and sample these regions only sparsely, they may by chance 
probe mostly galaxies with above average Ly$\alpha$ attenuation. We
find that this {\em sample variance} is not exceedingly large
for existing surveys, but that it does somewhat mitigate the required
neutral fraction at $z \sim 7$. Quantitatively, in a fiducial model
calibrated to match measurements after reionization, we find that
current $z=7$ observations require a volume-averaged neutral fraction
of $\avg{x_{\rm HI}} \geq 0.05$ at  95 per cent confidence level. Hence,
we find that the $z \sim 7$ Ly$\alpha$ fraction measurements do
likely probe the Universe before reionization completes but that they do
not require a very large neutral fraction.
\end{abstract}

\begin{keywords}
line: profiles -- galaxies: high redshift  --  intergalactic medium  --  
cosmology: theory  --  dark ages, reionization, first stars -- diffuse radiation.
\end{keywords}

\section{Introduction}

Recent studies have identified large samples of Ly$\alpha$ emitting galaxies (LAEs)
back to $z \sim 7$ (e.g. \citealt{Hu:2010um, Ouchi10, Kashikawa:2011dz}).
In addition to providing insight
into the properties of early galaxy populations, these surveys can be used to probe
the ionization state of the surrounding intergalactic medium (IGM) and the reionization
history of the Universe. In particular, the Ly$\alpha$ optical depth in a significantly neutral
IGM is so large that even the red side of a Ly$\alpha$ emission line should be 
attenuated by absorption in
the damping wing of the line \citep{MiraldaEscude:1997qb}. This is, in turn, expected to lead to a decline in the abundance
of observable LAEs as the Universe becomes significantly neutral. 
Indeed, \citet{Kashikawa:2006pb} found evidence for a decline in the abundance of LAEs between $z=5.7$ and
$6.5$ from observations in the Subaru Deep Field (SDF). Recent work has started to extend these measurements all the way
out to $z=7.7$ (e.g. \citealt{Clement12}); this study finds no LAE candidates, possibly indicating a further drop in the abundance
of LAEs from $z=6.5$ to $7.7$.

The declining LAE abundance may result if these observations are, in fact, probing
into the Epoch of Reionization (EoR) when the IGM is significantly neutral, but the existence of
a decline (e.g. \citealt{Hibon10, Hibon11, Hibon12, Tilvi10, Krug12}), its statistical significance
\citep{Ouchi10}
and the interpretation of the measurements (e.g. \citealt{Dijkstra:2006xx}) are all 
still debated.  One main challenge here is to
isolate the {\em part of the evolution in the LAE abundance that arises from
changes in the ionization state of the IGM}. The LAE abundance itself will, undoubtedly,
grow with time as a result of the hierarchical growth of the underlying LAE host halo population.
Furthermore, the {\em intrinsic properties} of the LAE populations should also evolve with time. Evolution
in the dust content, the structure of the interstellar medium, and the strength and  prevalence of large
scale outflows can all impact the escape of Ly$\alpha$ photons from galaxies, and the observable abundance
of the LAE populations.

In recent work, \citet{Stark10} proposed an approach that
partly circumvents concerns about intrinsic evolution in the underlying
galaxy populations. The \citet{Stark10} method uses the Lyman-break
selection technique to find populations of galaxies at the redshift of interest,
taking advantage of that technique's power in efficiently finding
many galaxies within a given redshift range. Then, follow-up spectroscopy
is done to determine which of those Lyman break galaxies (LBGs) have strong
enough Ly$\alpha$ emission to be classified as LAEs. Since reionization 
should produce little to no effect on the observation of LBGs, but Ly$\alpha$ emission is
attenuated by neutral hydrogen, the fraction of the LBGs that are also LAEs should
decrease as the IGM becomes significantly neutral. With this technique, evolution
in the abundance of the underlying population of star-forming galaxies `divides out'. Evolution
in the Ly$\alpha$ fraction induced by changes in the intrinsic LAE properties can be
extrapolated from lower redshift measurements of this fraction at $z \leq 5$ or so, which clearly probe the post-reionization epoch.
Moreover, the spectroscopic samples obtained in this approach allow one to search for spectroscopic
signatures of intrinsic evolution in the LAEs' properties. 

Several groups have recently applied this method for the first time; our focus here will be on the
theoretical interpretation of these new measurements. First, \citet{Stark10,Stark11} measured the
Ly$\alpha$ fraction ($f_{Ly\alpha}$)\footnote{Throughout we denote the Ly$\alpha$ fraction
by $f_{Ly\alpha}$, although some other works use $X_{Ly\alpha}$.} in several redshift bins, centred
around $z = 4, 5$, and $6$. The Universe should be completely ionized in this case, at least in the former two redshift bins, and
so this measurement describes the redshift evolution in the intrinsic LAE properties towards high redshift. These authors
find evidence that $f_{Ly\alpha}$ increases steadily from $z=4-6$.
They explain
this trend by noting that dust levels in these galaxies tend
to decrease with increasing redshift; this should likely facilitate
the escape of Ly$\alpha$ photons and so $f_{Ly\alpha}$ should grow
with
increasing redshift. Having
noted this trend, they make predictions for
\(f_{Ly\alpha}\) at \(z \sim\) 7, assuming there is no evolution in the
ionized fraction. A lower measurement of  \(f_{Ly\alpha}\)
than predicted would hence suggest that $z \sim 7$ observations are probing
the significantly neutral era. 

Very recently, several groups have extended these measurements out
to $z \sim 7$, including \citet{Schenker12}, \citet{Pentericci11},
\citet{Ono12} and \citet{Caruana12}. 
 As we discuss in more
detail subsequently, both \citet{Schenker12} and \citet{Pentericci11} see
evidence for a decline in $f_{Ly\alpha}$ near $z \sim 7$. Both Ono et al.'s and Caruana et al.'s results are consistent with little evolution from \citet{Stark11}'s lower redshift measurements
to within their error bars.
Although these measurements are still in their infancy, and come from $\leq 50$ LBGs in total, the
strong decline seems to suggest that the IGM is surprisingly neutral at $z \sim 7$.
In particular, \citet{Pentericci11} argue that the low $f_{Ly\alpha}$ indicated by
their $z \sim 7$ measurement requires a volume-averaged neutral fraction of
$\avg{x_{\rm HI}} \geq 0.6$. \citet{Schenker12} quote a fairly similar constraint, although somewhat more
tentatively.

Interestingly, the large neutral fractions suggested by this test are surprising given other constraints
on reionization. In particular, the relatively large optical depth to Thomson scattering implied by \emph{Wilkinson Microwave Anisotropy Probe} (\emph{WMAP}) observations \citep{Larson:2010gs} and
the low emissivity -- only a few ionizing photons per atom per Hubble time at $z \sim 5$ -- inferred from the mean transmission in the Ly$\alpha$ forest after reionization \citep{MiraldaEscude:2002yd,Bolton:2007fw} suggest that
reionization is a fairly prolonged and extended process.\footnote{However, a very recent analysis by \citet{Becker13}
favours a somewhat higher emissivity from the Ly$\alpha$ forest data. The higher inferred emissivity would help accommodate
more rapid reionization models.}
Furthermore, the absence of prominent Gunn--Peterson absorption troughs
in $z \leq 6$ quasar spectra \citep{Fan:2005es} is commonly taken to imply that the IGM is completely ionized by $z \leq 6$, a mere
$175$ Myr after the time period probed by the $z=7$ Ly$\alpha$ fraction measurements. Hence the Ly$\alpha$ fraction test results, when combined
with $z \leq 6$ quasar absorption spectra measurements, are hard to reconcile with other observational constraints -- as well as theoretical models for
the redshift evolution of the ionized fraction -- which all prefer a more extended reionization epoch.
One possibility is that reionization is sufficiently inhomogeneous
to allow transmission through the Ly$\alpha$ forest before the process completes \citep{Lidz:2007mz, Mesinger:2010mv}, potentially relaxing the requirement that reionization
needs to complete by $z \geq 6$ \citep{McGreer:2011dm}. Nevertheless, collectively current constraints on reionization prefer a significantly higher `mid-point' redshift at which 50 per cent of the volume of the IGM is neutral, $z_{50}$. For example, \citet{Kuhlen12} (see also \citealt{Pritchard:2010nm}) combine measurements of the
Thomson scattering optical depth, high-redshift LBG luminosity functions, and measurements from the post-reionization Ly$\alpha$ forest, finding
that the preferred redshift at which $\avg{x_{\rm HI}} = 0.5$ is $z_{50} = 10$ for a variety of models. 
 
However, as we look back into the EoR we expect both the average
abundance of LAEs to decrease and for there to {\em be increasingly
large spatial fluctuations in the abundance of observable LAEs} \citep{Furlanetto:2005ir,McQuinn07b,Mesinger:2008jr,Jensen:2012uk}. This should result from the patchiness of
the reionization process: LAEs that reside at the edge of ionized regions, or in small ionized bubbles,
will have significant damping wing absorption from nearby neutral hydrogen, while those at
the centre of large ionized bubbles will be less attenuated. Ultimately, measuring these spatial variations
is a very promising approach for isolating the impact of the IGM and studying reionization, with existing data already
providing some interesting constraints \citep{McQuinn07b,Ouchi10}. Presently,
the important point is that these fluctuations imply that one must survey a rather large region on the sky
to obtain a representative sample of the Ly$\alpha$ fraction. Perhaps existing $z \sim 7$ measurements
in fact probe later stages in reionization than previously inferred, but are sampling regions on the sky with above
average attenuation and correspondingly lower than average Ly$\alpha$ fractions. Here we set out to explore this sample variance effect, quantify its impact on
existing measurements, and to understand the requirements for future surveys to mitigate its impact.

The significance of the observed drops in the Ly$\alpha$
fraction also depends on the relationship between LAE and
LBGs; therefore, we examine a variety of models in an attempt to span some of the uncertainties
in the properties of the galaxies themselves. We use
simulations to explore the effects of small fields of view and
various models for LBG luminosity and Ly$\alpha$ emission on
\(f_{Ly\alpha}\) and, thus, what can be confidently said about
the ionized fraction of the Universe.

Throughout this paper, we assume a \(\Lambda\) cold dark matter  (\(\Lambda\)CDM) model with
\(n_{\rm{s}} = 1\), \(\sigma_{8} = 0.8\), \(\Omega_{\rm{m}} = 0.27\),
\(\Omega_{\Lambda}=0.73\), \(\Omega_{\rm{b}}=0.046\) and
\(h=0.7\). These parameters are broadly consistent with recent
results from the Planck collaboration \citep{Ade:2013zuv}.

\section{Observations}

First, we describe all of the current $z \sim 7$ Ly$\alpha$ fraction measurements. We subsequently compare these
measurements with theoretical models.
The main properties of the current observations are summarized in Table~\ref{fields}.

\citet{Pentericci11} describe observations they
obtained over three fields of view: the Great Observatories Origins Deep Surey (GOODS)-South
field \citep{Giavalisco04}, the New Technology
Telescope Deep Field (NTTDF; \citet{Arnouts99,
Fontana00}) and the BDF-4 field
\citep{Lehnert03}. Altogether, these fields span 
200 arcmin\(^2\) [roughly 700 (Mpc $h^{-1}$)$^2$]. Within these fields, they did
follow-up spectroscopy on galaxies that were selected
as $z$-dropouts. As we will discuss, in practice their spectroscopic sample probes the Ly$\alpha$ attenuation
across only a fraction of these full fields-of-view.
 Although these results are summarized
and combined by \citet{Pentericci11}, they also
published the initial measurements in stages. In
\citet{Fontana10}, they report their observations on
GOOD-S. In \citet{Vanzella11}, they do the same for
galaxies in BDF-4. Finally, in \citet{Pentericci11},
they discuss their observations of NTTDF. In these
fields combined, they found 20 LBGs, two of which had
strong enough Ly\(\alpha\) emission for them qualify
at LAEs.

They bin these galaxies based both on ultraviolet (UV)
magnitude and on the rest-frame equivalent widths of the Ly$\alpha$ lines, REWs. Since none of the
faint galaxies (\(-20.25<M_{UV}<-18.75\)) that
they observe has strong Ly$\alpha$
emission, for the faint case they are only
able to put upper limits on \(f_{Ly\alpha}\).
For the galaxies with stronger Ly$\alpha$
emission (REW \(>\) 55 \r{A}), their results
are consistent within error bars with the
projections from lower redshifts by
\citet{Stark11}. It is only for the weaker, but UV bright $(M_{\rm{UV}} < -20.25)$,
LAEs  (REW \(>\) 25 \r{A}) that they directly detect a
significant drop.

\citet{Schenker12} do spectroscopic follow-up
on LBGs with z \(>\) 6.3 observed in the \emph{Hubble Space Telescope (HST)}
Early Release Science (ERS) field by
\citet{Hathi10} and \citet{McLure11}, 36.5 arcmin$^2$ [119 (Mpc $h^{-1}$)$^2$].  In
addition to the eight galaxies they observed in
ERS, they also included 11 galaxies drawn from
a variety of other surveys. From this sample
of 19 galaxies, they found two that were
LAEs. In order to boost their sample size,
they also considered galaxies studied by
\citet{Fontana10} in their calculation of
\(f_{Ly\alpha}\).  Note that since
\citet{Fontana10} is part of  the sample sets
from both \citet{Pentericci11}  and
\citet{Schenker12}, these two sets of
observations are not completely
independent. \citet{Fontana10} detected seven LBGs
none of which was clearly LAEs. Thus,
\citet{Schenker12} worked from a total sample
of 26 LBGs, two of which they identified as
LAEs.

Combining these data sets and taking into
account the various limits of the
observations, they find that \(f_{Ly\alpha}\)
for the brighter galaxies
(\(-21.75<M_{\rm{UV}}<-20.25\)) is consistent with
the lower redshift measurements. It is only
for the faint galaxies
(\(-20.25<M_{UV}<-18.75\)) that they see a
significant drop. 

\citet{Ono12} observed the SDF and GOODS-N, a total of 1568
arcmin\(^2\) [roughly 5100 (Mpc $h^{-1}$)$^2$]. However, their survey is
much shallower than either \citet{Schenker12}
or \citet{Pentericci11}, only observing down
to \(y \simeq\) 26.0 mag
(\(M_{UV}=-20.9\)). Because of this shallowness,
even though they are observing a relatively
large area, they only identified 22 $z$-dropout
candidates \citep{Ouchi09}. From that sample,
they took spectroscopic observations of 11 of
them; three of those they identified as
LAEs. Their study only goes as deep as the
brightest of two \(M_{UV}\) bins
(\(-21.75<M_{UV}<-20.25\)). In that bin, their
results are consistent with the projections of
\citet{Stark11} from lower redshifts.

\citet{Caruana12} did spectroscopy for five $z$-band dropouts ($z \sim 7$) found in the \emph{Hubble Ultra Deep Field}, 11 arcmin$^2$ [36 (Mpc $h^{-1}$)$^2$], selected from earlier surveys. They observed no Ly$\alpha$ emission from the $z$-band dropouts, and, thus, only placed upper limits on the $f_{Ly\alpha}$ at $z \sim 7$; their upper limits are consistent with the projections from lower redshifts \citep{Stark11}.

One might wonder, for all of these surveys, about the presence of low redshift interlopers. If the fraction
of low redshift interlopers is larger near $z \sim 7$ than at lower redshifts, one might erroneously infer
a drop-off in the Ly$\alpha$ fraction. For instance, in the simplest version of the Lyman break selection technique
a lower redshift red galaxy may be mistaken for a higher redshift LBG. However, these groups have used a variety of techniques to ensure that the 
samples are truly from a redshift near \(z \sim 7\), preferring to discard those galaxies that are likely, but not definitively, at the redshift of interest in order to obtain an uncontaminated sample. In particular, \citet{Pentericci11}, as discussed in \citet{Castellano10}, use multiple filters and a number of colour selection criteria, to ensure that low-redshift interlopers are excluded. Similarly, the sample used by \citet{Ono12}, drawn from \citet{Ouchi09}, applies a variety of colour cuts to only select high redshift galaxies. \citet{Schenker12} derive their main sample from \citet{McLure11}. \citet{McLure11} fit the photometric redshifts of all potential candidates and required the ones that they retained to have both statistically acceptable solutions for $ z > 6 $ and to exclude lower redshift solutions at the 95 per cent confidence level. Based on all of these techniques, interloper contamination appears not to be a significant worry.

Since only \citet{Pentericci11} and
\citet{Schenker12} report significant
decreases in \(f_{Ly\alpha}\), the rest of the
paper will focus on their observations. As detailed
in Table~\ref{fields}, it is notable that the typical
comoving dimension of all but the shallower \citet{Ono12} survey
is $\sim 10$ Mpc $h^{-1}$. This scale is comparable to the size
of the ionized regions during much of the EoR (see e.g. \citealt{McQuinn07a}),
suggesting that sample variance may be significant on these scales. Furthermore, as we detail subsequently, the 
LBGs identified for follow-up spectroscopy in these fields only {\em sparsely sample} sub regions of the full field,
and so the surveys do not, in fact, sample Ly$\alpha$ attenuation across the entire field of view.

\begin{table*}
\begin{minipage}[c]{176mm}
\caption{Summary of published measurements of
\(f_{Ly\alpha}\)}
\begin{center}
\begin{tabular}{ r r r r r r r r r}
\hline \multirow{2}{*}{Field} &
\multirow{2}{*}{Area}
&\multirow{2}{*}{Shortest side} &
\multirow{2}{1.6cm}{Limiting mag ($y$
  band)}
&\multicolumn{2}{c}{Faint\(^1\)}
&\multicolumn{2}{c}{Bright\(^b\)}
&\multirow{2}{*}{Source\(^c\)}\\ \cline{5-8}
& & & &  LBGs & LAEs & LBGs & LAEs &
\\ \hline \hline
\multicolumn{9}{c}{\citet{Pentericci11}}\\ \hline
\hline BDF-4& \(\sim\) 50
arcmin\(^2\)&13.6 Mpc $h^{-1}$&26.5& 1 & 0 &
5&2&\citet{Vanzella11,
  Pentericci11}\\ \hline GOODS-S&
\(\sim\) 90 arcmin\(^2\)& 13.6 Mpc $h^{-1}$&
26.7& 1 & 0
&6&0&\citet{Fontana10}\\ \hline NTTDF
&\(\sim\) 50 arcmin\(^2\)& 13.6 Mpc $h^{-1}$
&  26.5 & 0 & 0 & 7
&0&\citet{Pentericci11}\\ \hline
\hline
\multicolumn{9}{c}{\citet{Schenker12}}\\ \hline
\hline GOODS-S& \(\sim\) 90
arcmin\(^2\)& 13.6 Mpc $h^{-1}$ &26.7& 1 & 0
&6&0&\citet{Fontana10}\\ \hline ERS &
36.5 arcmin\(^2\) &7.7 Mpc $h^{-1}$& 27.26& 8
& 0 & 0 & 0
&\citet{Schenker12}\\ \hline Other
fields\(^c\) & -- & -- & -- & 7 & 2 & 4 &
0 &\citet{Schenker12}\\ \hline \hline
\multicolumn{9}{c}{\citet{Ono12}}\\ \hline
\hline GOODS-N& \(\sim\) 150
arcmin\(^2\)& 18.1 Mpc $h^{-1}$&26.0 & 0 &0&1
&1&\citet{Ono12}\\ \hline SDF&
\(\sim\) 1400 arcmin\(^2\) &48.9 Mpc $h^{-1}$
& 26.0& 0 &0&7
&2&\citet{Ono12}\\ \hline \hline
\multicolumn{9}{c}{\citet{Caruana12}}\\ \hline
\hline HUDF& \(\sim\) 11
arcmin\(^2\)& 6.0 Mpc $h^{-1}$& --\(^e\)& 3 &0&2
&0&\citet{Caruana12}\\ 
\hline
\multicolumn{9}{l}{\footnotesize{\(^a\)Faint
    galaxies are ones with
    \(-20.25<M_{UV}<-18.75\)}}\\ \multicolumn{9}{l}{\footnotesize{\(^b\)Bright
    galaxies are ones with
    \(-21.75<M_{UV}<-20.25\)}}\\ \multicolumn{9}{l}{\footnotesize{\(^c\)May
    not have originally observed these
    galaxies,  but did the
    spectroscopic follow up and
    reported which ones had
    Ly\(\alpha\) emission
}}\\ \multicolumn{9}{p{16cm}}{\footnotesize{\(^d\)
    \citet{Schenker12} included
    several galaxies from a variety of
    other surveys and fields in their
    spectroscopic sample. These have been grouped together here. }}\\
\multicolumn{9}{p{16cm}}{\footnotesize{\(^e\) \cite{Caruana12} did spectroscopic follow up on galaxies drawn from several surveys. Thus, there is no consistent y-band limiting magnitude.}}\\
\multicolumn{9}{p{16cm}}{Summary of
  observations for the main
  papers we are discussing. The
  \citet{Fontana10} sample is included in
  both the observations of
  \citet{Pentericci11} and
  \citet{Schenker12}. This is because
  both groups used that data in their
  calculation of
  \(f_{Ly\alpha}\). 
Galaxies were only counted as LAEs if their Ly$\alpha$
emission was greater than some minimum threshold (usually
an equivalent width threshold of 25 \r{A}).
In compiling this table,
  summaries in \citet{Ono12} were
  essential.}

\end{tabular}
\end{center}
\label{fields}
\end{minipage}
\end{table*}

\section{Method}

Next we describe the various ingredients that enter into our simulated models of
the Ly$\alpha$ fraction. Our calculations start from the reionization simulations
of \citet{McQuinn07b}; these simulations provide realizations of the inhomogeneous ionization field and surrounding
neutral gas, which in turn modulate the observable abundance of the simulated LAEs. In post-processing steps, we further
populate simulated dark matter haloes with LBGs and LAEs, finally attenuating the LAEs based on the simulated neutral
hydrogen distribution. A completely self-consistent approach would vary the reionization history jointly with variations in
the properties of the LBGs and LAEs, but our post-processing approach offers far greater flexibility for exploring a wide
range of models. In addition, most of the ionizing photons are likely generated by much fainter, yet more abundant, sources than the LBGs
that are detected directly (thus far) and so the ionization history is somewhat decoupled from the LBG properties relevant here.
Our philosophy for modelling the LAEs is to focus our attention on the impact of reionization and its spatial inhomogeneity.
As a result, we presently ignore the complexities of the Ly$\alpha$ line transfer internal to the galaxies themselves; we will comment
on their possible impact on our main conclusions.

\subsection{Reionization simulations}

The reionization simulations used in our analysis are described in \citet{McQuinn07a,McQuinn07b}; here we mention
only a few pertinent details. These simulations start from a $1024^3$ particle, $130$ comoving Mpc $h^{-1}$ dark matter simulation
run with \textsc{gadget-2} 
\citep{Springel05} and treat the radiative transfer of ionizing photons in a post-processing step. 
The calculation resolves host haloes down to $\sim 10^{10} M \odot$, and so the simulation
directly captures the likely host haloes of both LBGs and LAEs (smaller mass haloes host ionizing sources and 
are added into the simulation as described in \citet{McQuinn07b}). We adopt the fiducial model of these authors in which each halo above
the atomic cooling mass (on the order of $M \sim 10^8 M \odot$, \citealt{Barkana01}) hosts a source with an ionizing luminosity that is directly proportional to its halo mass.
We consider simulation outputs with a range of ionized fractions, $\avg{x_i}$, in effort to explore how the Ly$\alpha$ fraction and
its spatial variations evolve throughout reionization. In each case, we take the host haloes from a simulation output at $z=6.9$, very close
to the redshift of interest for the current Ly$\alpha$ fraction measurements. In the fiducial model of \citet{McQuinn07b} studied here, the volume-averaged
ionization fraction is $\avg{x_i}=0.82$ at this redshift. In practice, when we explore smaller ionized fractions, we use slightly
higher redshift outputs of the ionization field in the same model. This approximation was also adopted in \citet{McQuinn07b} and should be
adequate since -- at fixed $\avg{x_i}$ -- the properties of the ionization field depend little on the precise redshift at which the ionized
fraction is reached (tests of this are given in \citealt{McQuinn07a}).

\subsection{LBG model}

The next step in our model is to populate the simulated
dark matter haloes with LBGs. Here we follow the simple 
approach described in \citet{Stark07}. In this model, 
the star formation
rate, $\dot{M}_\star$, is connected to the halo mass, $M_{\rm halo}$, by
\begin{equation}
\dot{M}_\star = \frac{f_\star (\Omega_{\rm{b}}/\Omega_{\rm{m}}) M_{\rm{halo}}} {t_{\rm{LT}}},
\label{eq:mdot_star}
\end{equation}
where $f_\star$ is the efficiency at which baryons are converted into stars
and $t_{\rm{LT}}$ is the time-scale for star formation, which is itself the product
of the star formation duty cycle ($\epsilon_{\rm{DC}}$) and the age of the Universe which -- in 
the high-$z$ approximation -- is $t(z) = 2/(3H(z))$. We assume here that stars form at a constant
rate in the haloes that are actively forming stars. 
We adopt \citet{Stark07}'s best-fitting values of \(f_{*}=0.16\) and \(\epsilon_{\rm{DC}}=0.25\),
for LBGs at \(z\simeq 6 \). We further assume the conversion from star formation
rate to UV luminosity (at a rest-frame wavelength of $\lambda = 1500 \Ang$)  given in \citet{Madau98}:
\begin{equation}
L_{1500} = 8.0 \times 10^{27} \frac{\rm{erg}}{\rm{Hz\ s}} \left[\frac{\dot{M}_\star}{1 M \odot/\rm{yr}}\right].
\label{eq:luv}
\end{equation}
The conversion factor here assumes a Saltpeter initial mass function and solar metallicity, but we do not expect our results
to be sensitive to the precise number here. In the simple model described here, $\epsilon_{\rm{DC}}$ also gives the fraction
of host haloes that are UV luminous at any given time. However, we find that the observed number density of LBGs at $z \sim 7$ from
\citet{Bouwens11} is better reproduced if instead $1/6$th of simulated dark matter haloes host LBGs with the above luminosity.  In practice
we randomly select $1/6$th of our dark matter haloes as active LBG hosts and give them UV luminosities according to the above formulas 
(with $\epsilon_{\rm{DC}} = 0.25$). This also allows us to reproduce the bias factor found from large samples of LBGs \citep{Overzier06}. 
The assumption here of a one-to-one relation between UV luminosity and host halo mass for the active hosts is undoubtedly a simplification, but
we have experimented with adding a $20$ per cent scatter to the UV luminosity--halo mass relation above and found little difference in our results.

\subsection{Properties of the LAEs}
\label{sec:LAE_props}

Next, we give each LBG a random amount of Ly$\alpha$ emission, as characterized by the REW of each emission line.
Our main interest is to explore the impact of inhomogeneous damping-wing absorption on the Ly$\alpha$ fraction,
and so we do not attempt a detailed modelling of each emission line. In addition to the complex impact of the Ly$\alpha$ line
transfer internal to each galaxy, the resulting line will be modified by resonant absorption within the IGM (in addition to the
damping wing absorption that we do explicitly model). It is sometimes assumed that each emission line is symmetric upon leaving the
galaxy, and that the blue side of each line is removed by resonant absorption in the surrounding IGM while the red side
is fully transmitted. The reality is of course more complicated, and even the resonant absorption in the IGM may have a wide
range of effects depending on the strength of large-scale infall motions towards LAE host haloes (which can lead to resonant
absorption on the red side of the line), the local photoionization state of the gas, outflows, and other factors (see e.g.
\citealt{Santos:2003pc,Dijkstra:2007nj,Zheng:2009ax,Verhamme12,Yajima12,Duval13}). 
We will assume that these variations are captured by drawing the REW of each LBG from a random distribution. We will refer to this
loosely as the {\em intrinsic} REW, and that the observed REW results from simply multiplying this intrinsic REW by a damping wing
attenuation factor described below. In this sense, our intrinsic REW is intended to include resonant absorption from the IGM, and
we implicitly assume that this distribution does not vary strongly from $z \sim 6$ to $z \sim 7$. We additionally neglect any
spatial variations -- induced by e.g., large scale infall motions -- in the intrinsic REW distribution.

As in \citet{Dijkstra11}, we draw an REW for each simulated galaxy from an exponential distribution, $P(\rm{REW}) = \rm{exp}[-REW/REW_c]/REW_c$.
This exponential model provides a good fit to the distribution of REWs found in lower redshift galaxies \citep{Gronwall07, Blanc11}.
We further allow the REW to be either uncorrelated, correlated, or anti-correlated with the UV luminosity of each simulated LBG. In our
fiducial model, the REW and UV luminosity are anti correlated since this trend appears to best reproduce existing observations. 
In order to produce REW distributions that are anti correlated with UV luminosity, we draw REWs from the exponential distribution (with
$\rm{REW_c} = 125 \Ang$ in our fiducial model), and
place them into various REW bins that are then populated with LBGs according to their UV luminosity. Specifically, we divide REWs
into 10 bins based on width: the
first bin is from 0 to 32 \r{A}, the second
from 32 to 110 \r{A} and the remaining eight
are 80 \r{A} wide. The LBGs are then divided into 10 bins based on UV luminosity, and the most luminous galaxies are randomly assigned
REWs from the first (smallest REW) REW bin, the next most luminous from the first and second bins, and so forth, until all the LBGs
have REWs. Although the details of this procedure are somewhat arbitrary, it matches both the observed values of $f_{Ly\alpha}$ at
$z \sim 6$ (for both UV bright and UV faint sources), as well as the correlation coefficient, $\rho$, between REW and UV luminosity. 
This is defined as
\begin{equation}
	\rho=\frac{\sum\limits_{i=1}^n (\rm{REW_i}-\overline{\rm{REW}})(L_i-\overline{L})}{\sqrt{\sum\limits_{i=1}^n (\rm{REW}_i-\overline{\rm{REW}})^2} \sqrt{\sum\limits_{i=1}^n (L_i-\overline{L})^2}}.
	\label{eq:rho}
\end{equation}
The sum here is over the LAEs in the observed or simulated sample, with $n$ LAEs in total, $\rm{REW_i}$ denotes the rest frame
equivalent width of the $i$th LAE in the sample, $L_i$ is the UV luminosity of the $i$th LAE, while $\overline{\rm{REW}}$ and
$\overline{L}$ denote the sample-averaged rest frame equivalent width and UV luminosity, respectively.  
The correlation coefficient in our fiducial model is $\rho = -0.23$, while from fig. 12 of \citet{Stark10} we infer
$\rho= - 0.41$; since the variance of the observational estimate is likely still sizeable, we consider this simple model to be
broadly consistent with observations.

This negative correlation model is also physically plausible, as discussed in \S \ref{sec:model_dep}.
Of course, our
fiducial description of this correlation is only a toy model. Nevertheless, we believe that it captures
the relevant features of the lower redshift observations, and, thus, should be
a useful guide as we consider the $f_{Ly\alpha}$ measurements.

In addition to our fiducial model, we discuss
three other models that seem both physically
plausible and show a diversity of results: (1)
the REWs are assigned randomly; (2) there is a
weak positive correlation between the REW and
continuum luminosity of the galaxy and (3)
there is a weak negative correlation between
the REW and the continuum luminosity of the
galaxy -- weaker than in our fiducial case. For the second and third cases, the
REWs are assigned in a fashion similar to that
in our fiducial case, except the galaxies, and
REWs, are only divided into three bins, not
10. Thus, the correlation is weaker than in
the fiducial case. For the weak positive
correlation, the correlation coefficient is
$\rho = 0.10$;  for the weak negative correlation,
$\rho = -0.11$. While these models are unlikely to
perfectly capture the relationship between LAEs and LBGs, they
suffice to explore how our results depend on the precise
relationship between LBGs and LAEs.

\subsection{Attenuated LAE emission and mock surveys}

The intrinsic emission from each LAE is then attenuated according to
the simulated distribution of neutral hydrogen. We do this following
\citet{McQuinn07b} and \citet{Mesinger:2008jr}.
Specifically, we shoot
lines of sight through the simulation box towards each LAE and calculate
the total damping wing contribution to the Ly$\alpha$ optical depth, summing over
intervening neutral patches. As in this
previous work, we calculate the damping wing optical depth $\tau_{\rm{DW}}$ (according to e.g.,
equation 1 of \citealt{Mesinger:2008jr}) only at the centre of each emission line, i.e. at
an observed wavelength of $\lambda_{\rm{obs}} = \lambda_\alpha (1+z_{\rm{s}})$ for a source at
redshift $z_{\rm{s}}$, with $\lambda_\alpha = 1215.67 \Ang$ denoting the rest-frame line centre
of the Ly$\alpha$ line. We ignore the impact of peculiar velocities and neglect any redshift
evolution in the properties of the ionization field across our simulation box -- we use
snapshots of the simulated ionization field at fixed redshift.
 Calculating the
optical depth at the centre of each line, rather than the full profile, should be a good approximate
indicator of which sources will be attenuated below observational REW cut-offs (see \citealt{McQuinn07b} for a
discussion and tests). The `observed' REW of each simulated LBG is then an attenuated version of the initial intrinsic REW according to 
$\rm{REW}_{\rm{obs}} = \rm{e}^{-\tau_{\rm{DW}}} \rm{REW}$.

Armed with galaxy positions, UV luminosities and
observed REWs, we produce mock realizations of the
\citet{Schenker12} and
\citet{Pentericci11} observations, and measure their statistical properties.
We slice the simulation
cube into strips with the perpendicular
dimensions mimicking the geometry and field
of view of the observations. In order to approximately
mimic the survey window functions in 
the redshift direction, we assume each strip has
a length of 100 comoving Mpc $h^{-1}$. We adopt this
value based on the estimated redshift distribution
of $z$'-band dropouts in \citet{Ouchi09} (their fig. 6), which those authors determined 
from the SDF and GOODS-N fields. For simplicity, we
assume a top-hat window function and find that a top-hat of width 100 Mpc $h^{-1}$
reproduces the area under the redshift distribution curve of
\citet{Ouchi09} and so use this width throughout.  While all three of these surveys \citep{Ouchi09, Pentericci11, Schenker12} use slightly different filters, they all peak at roughly the same wavelength and \citet{Ouchi09}'s have the largest full width at half maximum (FWHM). Thus, this should be a reasonable approximation, although it may slightly overestimate the depth of the field.

Within these strips, we then select the
galaxies that would be observable to
the two groups, based on the limiting
magnitudes and REW limits they reported (see Table~\ref{fields}).
One subtlety here is that the galaxies observed
by both groups are not distributed across
their entire fields. This likely occurs partly because
the LBGs are clustered, and partly because their efficiency
for detecting LBGs and performing spectroscopic follow-up
is not uniform across each field. Because of the latter effect,
we may underestimate the sample variance if we draw the
simulated LBGs from the full field of view. To roughly mimic
this, we draw our LBGs from randomly placed cylinders with diameters
matching the largest separation between galaxies in each set of
observations: 9.99 Mpc $h^{-1}$ for \citet{Schenker12} and 8.04 Mpc $h^{-1}$ for \citet{Pentericci11}. We only consider observable galaxies that fall within each cylinder,
effectively limiting the size of the observed fields and making them still closer
to the size of the ionized bubbles in the simulation. 

Furthermore, comparing to the $z \sim 7$ LBG luminosity function from \citep{Bouwens11}, it is clear that the
existing measurements perform spectroscopic follow-up on only a small fraction of the total number of LBGs expected in each survey field. 
This likely owes mostly to the expense of the
spectroscopic follow-up observations. However, this implies that the Ly$\alpha$ fraction
measurements only sparsely sample the Ly$\alpha$ attenuation across the entire field-of-view, and enhances the sample variance effect. To mimic
this, in each mock survey we randomly select LBGs to match the precise number spectroscopically observed, and consider the Ly$\alpha$ emission and attenuation
around only these galaxies. In the case of \citet{Schenker12}'s ERS field, they actually do not observe any bright LBGs. To loosely consider this case -- which
will turn out to be less constraining -- we randomly select four simulated LBGs, which matches their bright sample in the other fields and also corresponds
to the number one would expect based on the size of their field of view and the follow-up capabilities in \cite{Pentericci11}. 

One caveat here is that our simulation does not capture Fourier modes that are larger than our simulation box size (130 comoving Mpc $h^{-1}$), and so the sample variance we estimate is only a lower bound. We do not expect the `missing variance'
to be large, however. Although the full fields of view of some of the existing surveys correspond to a decent fraction
of our simulation volume (see Table~\ref{fields}), the effective volume probed by the observations 
is significantly smaller, as a result of the sparse sampling discussed previously. In addition, it is important
to note that our simulation box size is large compared to the size of the ionized regions during most of the
reionization epoch, and so it does suffice to capture a representative sample of the ionization field.

Since both teams did follow up
spectroscopy on already identified LBGs, we
look for Ly$\alpha$ emission in the
selected LBGs that is strong enough
to exceed the observed REW cuts. The field-to-field fluctuations in $f_{Ly\alpha}$ arise from both the sample variance we aim to study and from discreteness
fluctuations. In our model, the latter arise because the REW of the Ly$\alpha$ emission from each LBG is drawn from a random distribution. This scatter
is, of course, non-vanishing even in the absence of reionization-induced inhomogeneities, and moreover, it is already included in the error budget of
the existing measurements (while the sample variance contribution has not been included). Hence we must separate the discreteness and sample variance contributions
to the simulated variance in order to avoid double counting the discreteness component. To do this, we keep track of which LBGs fluctuate above and below the observational
REW cuts as a result of an upward or downward fluctuation in the IGM attenuation. Some LBGs in the simulated sample are observable as LAEs only because of downward fluctuations in the IGM attenuation, while some are pulled out of the LAE sample because of upward fluctuations in the IGM attenuation. Finally, in some cases -- depending on the LBG's intrinsic REW and IGM attenuation -- the IGM attenuation does not impact the LBG's observability as an LAE. Monitoring
which LBGs are pulled into or out of the sample owing to fluctuations in the IGM attenuation allows us to measure the scatter in $f_{Ly\alpha}$ arising from
sample variance alone. 

\citet{Schenker12} combine observations from a variety of surveys. Their observations focus on galaxies selected from an LBG survey on the ERS field. To increase their sample size, \citet{Schenker12} include observations from \citet{Fontana10} and several other LBGs from a variety of fields in their analysis. Since \citet{Fontana10} is included in \citet{Pentericci11}'s analysis and the other LBGs were not all detected in well defined drop-out surveys of the corresponding fields, we only model the observations of the ERS field of view when we analyse Schenker's results.

\citet{Pentericci11} observed with four
pointings of the High Acuity Wide Field K-Band Imager (HAWK-I;  two in GOODS-S, one
at BDF-4 and one at  NTTDF). Although the two
fields in GOOD-S are next to each other, for
ease of calculation we treated each pointing
as independent, and scaled the confidence
regions accordingly. This may lead us to underestimate
the sample variance from these observations slightly.

\section{Discussion}

\subsection{Is a large neutral fraction required?}

\begin{figure*}

        \begin{center}

                \includegraphics[scale=.3]{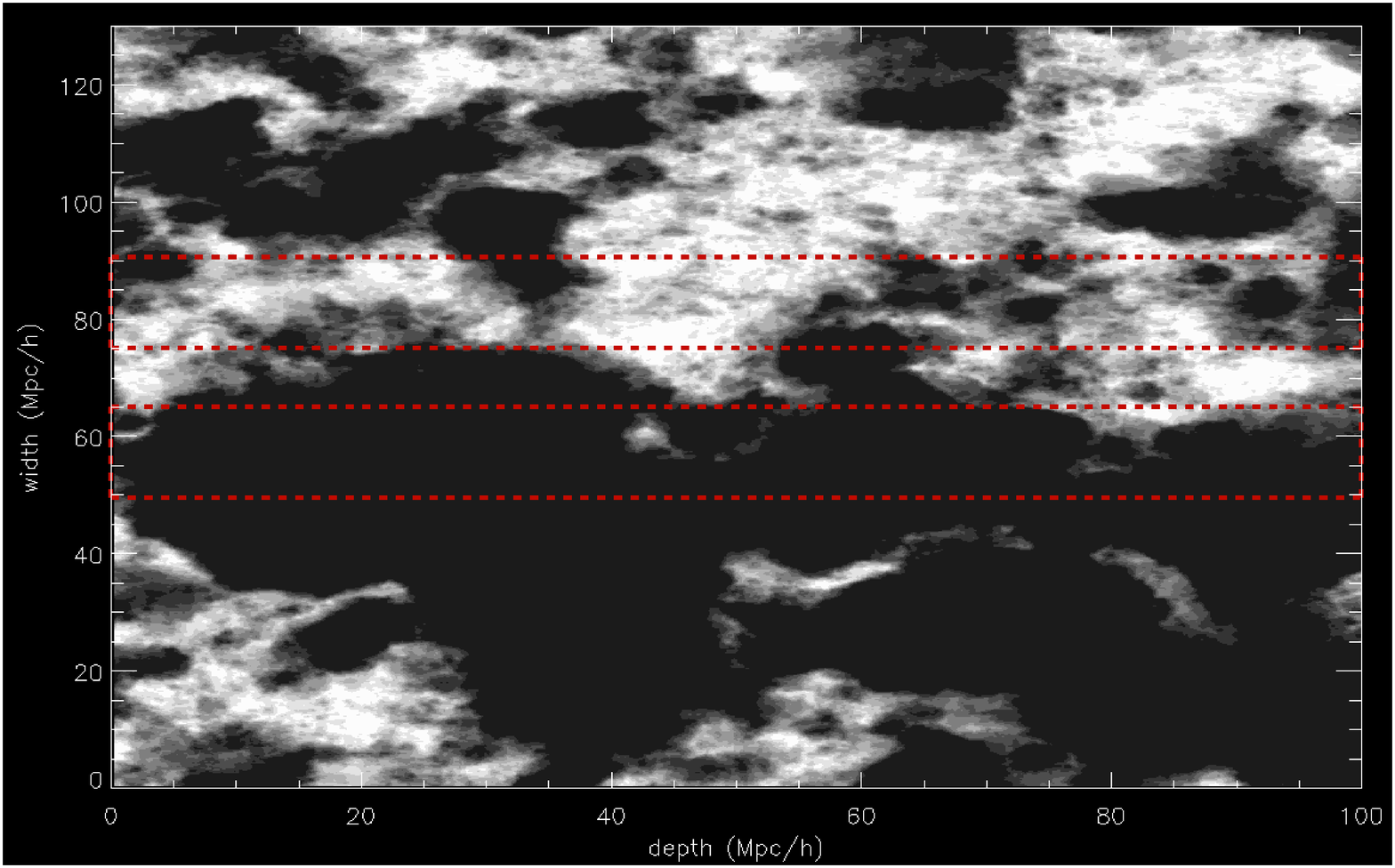}

                \caption{{A slice through the
                    simulation cube for an output with a
                    volume-averaged ionization
                    fraction of \(\avg{x_{\rm{i}}}=
                    0.82\). The slice is
                    averaged over one (perpendicular to the line of sight)
                    dimension of the ERS field
                    of view. Clouds of neutral
                    gas are shown in
                    grey-scale; the white
                    areas being completely
                    neutral. The
                    dashed lines mark
                    two potential mock ERS
                    observations. 
The significant difference between these two fields illustrates
that small regions on the sky may not provide representative samples of the overall average ionization state of the IGM.}}
                \label{slices}

        \end{center}

\end{figure*}

Before we present a detailed comparison between the Ly$\alpha$ fraction measurements
and our simulated models and explore their implications for understanding
reionization, a qualitative illustration of the main effect discussed here may be helpful. This is provided by Fig. ~\ref{slices}, which shows a representative slice through the
simulation volume, when the volume-averaged ionization fraction is
$\avg{x_{\rm{i}}} = 0.82$. The dimension into and out of the page has been averaged
over the size of an ERS field. The two red regions show example
mock survey volumes that have the same size as an ERS field. Clearly the top field is fairly well covered by intervening neutral hydrogen, although it is still has patches of ionized hydrogen. On the other hand, the bottom cylinder is almost entirely clear
of neutral hydrogen. LBGs residing in the top cylinder will be much more
attenuated than in the bottom panel. 

Quantitatively, in the faint UV luminosity bin of \citet{Schenker12}, the simulated $f_{Ly\alpha}$ in 
the first simulated field is $f_{Ly\alpha} = 0.30$, while the Ly$\alpha$ fraction is almost twice as large in the second simulated field, 
$f_{Ly\alpha} = 0.58$. These numbers can be compared to the average across the entire simulation
volume at this neutral fraction, which is $f_{Ly\alpha} = 0.51$. Using our fiducial model and supposing (incorrectly) that these fractions are representative, one would infer an ionized fraction of $\avg{x_{\rm{i}}} = 0.54$ for the first field, compared to $\avg{x_{\rm{i}}} = 1.0$ for the second field. Neither inference correctly returns the true ionized fraction of the field, $\avg{x_{\rm{i}}} = 0.82$. This suggests that a rather large area survey is indeed required to obtain representative measurements of the Ly$\alpha$ fraction. Perhaps
the existing fields are more like the top cylinder, and the LBGs in these fields suffer above average attenuation. Assuming a survey like the top
cylinder is representative of the volume averaged neutral fraction could clearly bias one's inferences of the neutral fraction. The chance of obtaining a biased estimate of the neutral fraction is enhanced if spectroscopic follow-up is done on only a few LBGs -- effectively, a sparse sampling of each survey region -- as is often the case for the existing measurements.

With this intuitive understanding, we turn to a more detailed comparison with the
observations. \citet{Schenker12} report a
decline in \(f_{Ly\alpha}\) at $z$ \(\sim\)
7. This decline is particularly pronounced
when compared to their observed 
\(f_{Ly\alpha}\) values at lower redshifts. For
\(4<z<6\), they report an increase in the
Ly$\alpha$ fraction with increasing redshift
\citep{Stark11}. At $z$ \(\sim\) 7,
\(f_{Ly\alpha}\) is lower than the projected
trend and lower than measured at $z$ \(\sim\) 6. This
decline suggests that that the observations may be
probing into the EoR. Using Monte Carlo simulations of
their full data sample, \citet{Schenker12} conclude
that in order to explain their results they
require an ionized hydrogen fraction of at most
$\avg{x_{\rm{i}}} \leq 0.51$. 

Note that the low redshift measurements (\(4<z<6\)) come mainly
from observations of the GOODS fields, both North and
South. Some additional galaxies from other sources are
also included. Altogether, \citet{Stark11} observed
351 $B$-dropouts, 151 $V$-dropouts, and 67
\textit{i'}-dropouts. Given the large sky coverage of these
surveys, sample variance should not be a significant source of
error for the low-redshift measurements. However, the same
may not be true at higher redshift where the measurements come
from smaller fields of view. In addition, if the high-redshift
measurements probe into the EoR, this should enhance the sample variance
as we will describe.

\begin{figure}

        \begin{center}

          \includegraphics[scale=.30]{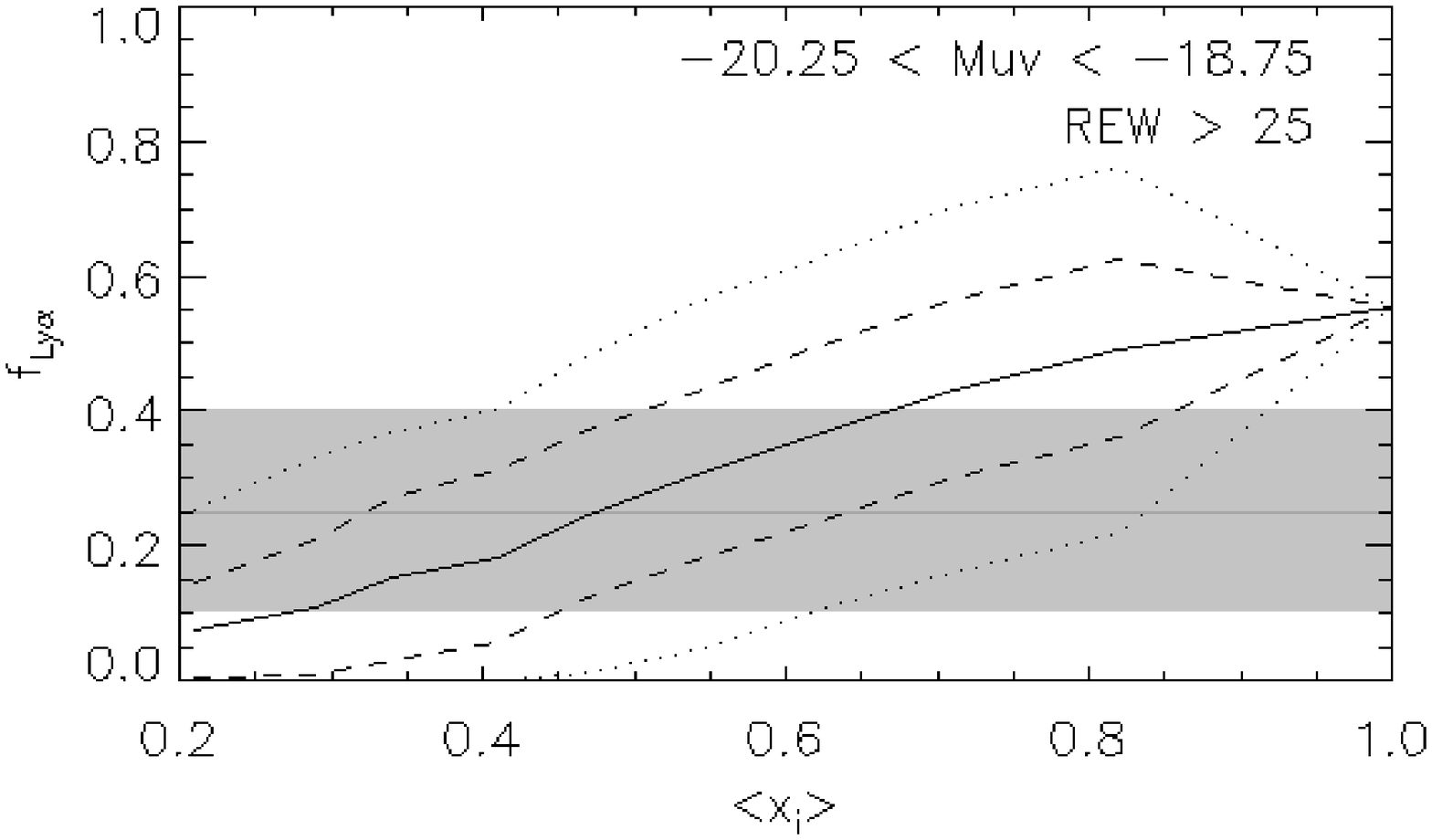}

          \includegraphics[scale=0.30]{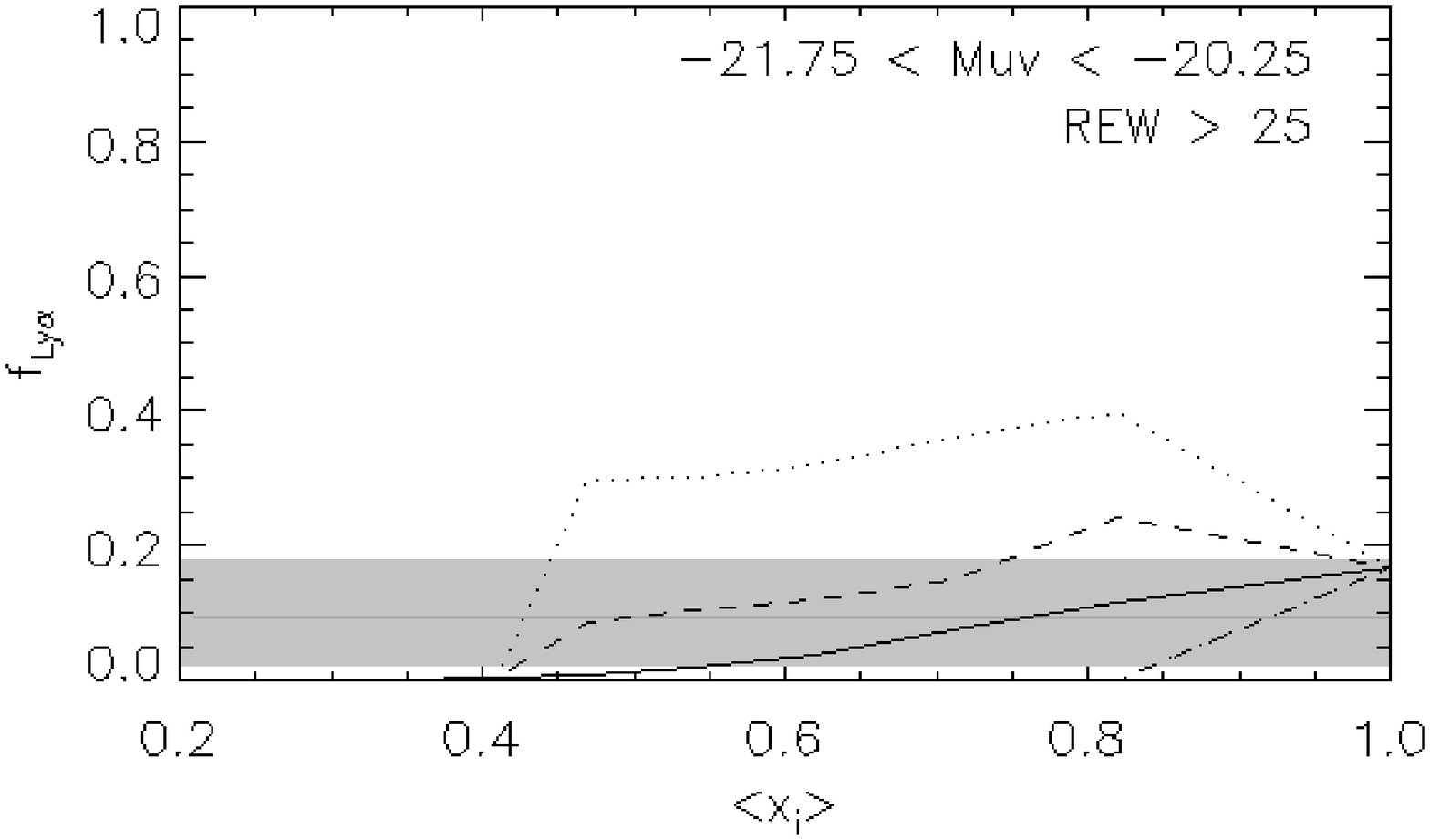}

          \caption{Comparison between mock Ly$\alpha$ fraction ($f_{Ly\alpha}$) measurements and
the observations of \citet{Schenker12}, as a function of the volume-averaged
ionization fraction, $\avg{x_{\rm{i}}}$.  The solid lines show the average value of $f_{Ly\alpha}$ across the simulation volume in our fiducial model, while the dotted and dashed lines indicate the field-to-field spread in the simulated Ly$\alpha$ fraction, i.e. the sample variance. The dashed lines enclose the values of
$f_{Ly\alpha}$ for 68 per cent of the mock ERS fields, while the dotted lines enclose 95 per cent of the simulated fields. The shaded regions show the 68 per cent confidence interval
reported by \citet{Schenker12} (which neglect the sample variance contribution). The horizontal line near the centre of the shaded
bands gives their best-fitting value for the Ly$\alpha$ fraction.
 The top panel is for the UV faint bin
of \citet{Schenker12}, while the bottom panel is for the UV bright bin.}

                \label{fig:ERS}

        \end{center}

\end{figure}	

In Fig. \ref{fig:ERS} we show simulated Ly$\alpha$
fractions for mock surveys mimicking the $z \sim 7$ measurements of
\citet{Schenker12} for the ERS field. %As noted above, the other LBGs they include in their survey, are either part of Pentericci's survey and are, thus, included in this work when we discuss \citet{Pentericci11} or are not part of a systematic LBG survey so we do not include them.
In particular, we plot the average Ly$\alpha$
fraction across the entire simulation box (solid line), as well as 68 per cent (dotted lines) and
95 per cent (dashed lines) confidence intervals, as a function of the volume-averaged
ionization fraction ($\avg{x_{\rm{i}}}$) in the models. 
The lines reflect the field-to-field variance (i.e. the sample variance) in the
simulated $f_{Ly\alpha}$ values. The probability distribution of simulated $f_{Ly\alpha}$ values
is somewhat non-Gaussian, and so we account for this in determining the confidence intervals.
In order to compare with \citet{Schenker12} we divide the simulated LBGs into UV bright ($-21.75 < M_{UV} < -20.25$) (bottom panel)
and UV faint ($-20.25 < M_{UV} < -18.75$) (top panel) bins. The shaded region in each panel gives the allowed range in $f_{Ly\alpha}$ at
68 per cent confidence reported by \citet{Schenker12}, while the horizontal line shows their best-fitting $f_{Ly\alpha}$ value. In calculating the 68 per cent confidence range, \citet{Schenker12} include LBGs from several fields. As discussed earlier, we only calculate the sample variance for the ERS field. The sample variance may thus be slightly overestimated when compared to the rest of the error budget.  The shaded
region neglects the sample variance contribution, and so a better estimate of the total error budget is the quadrature sum of the shaded regions
and the dashed lines.

Focusing first on the UV faint bin (top panel), the first conclusion we draw from this comparison
is that the {\em best-fitting} ionized fraction is indeed $\avg{x_{\rm{i}}} \approx 0.5$. This is consistent with the conclusions of
\citet{Schenker12}. Note, however, that \citet{Schenker12} infer the neutral fraction from the same \citet{McQuinn07b} simulations used here,
and so their constraint is not independent of the results here, although our present calculations adopt a different model for the LAE/LBG populations.
The next conclusion to draw from the top panel is that the spread from sample variance is generally comparable to (although a little smaller than)
the reported error budget, and so it is not in fact negligible, aside from the fully ionized case where it vanishes in this model. Here the sample variance
describes spatial variations in the damping wing attenuation, which vanish in the fully ionized model. In reality, spatial variations in the resonant
absorption likely lead to additional contributions, neglected here, which would make the sample variance non-vanishing in the fully ionized case.
Accounting for sample variance somewhat reduces the neutral fraction required by these observations; for instance an $\avg{x_{\rm{i}}} \approx 0.85$ model
is just outside of the allowed 68 per cent confidence range.

The bottom panel shows that the UV bright case is still easier to accommodate in a highly ionized IGM, at least in our fiducial LAE model. 
In this case, \citet{Schenker12} found little evidence for redshift evolution. Since our fiducial model is constructed to give a lower
$f_{Ly\alpha}$ fraction for UV bright LBGs after reionization, the Ly$\alpha$ fraction remains small at $z \sim 7$ in a highly ionized IGM.
In fact, since the UV bright galaxies tend to have smaller intrinsic REWs in our fiducial model, these galaxies are particularly susceptible
to attenuation: that \citet{Schenker12} detect some Ly$\alpha$ emission from UV bright galaxies in this model then actually {\em slightly disfavours}
more neutral models. At the 68 per cent confidence level,  $\avg{x_{\rm{i}}} \approx 0.42$ or smaller is disfavoured. The relatively small number of LBGs spectroscopically followed up in each field (only four) contributes to the size of the error bars. Recall that our sampling of four model galaxies here is somewhat arbitrary, as \citet{Schenker12} did not in fact follow up any bright LBGs in the ERS field.
 Further, the trend of LAE fraction with ionization fraction depends somewhat on our particular model for the anti-correlation between intrinsic REW
and UV luminosity, as we describe in \S \ref{sec:model_dep}.

\begin{figure}

        \begin{center}

          \includegraphics[scale=0.30]{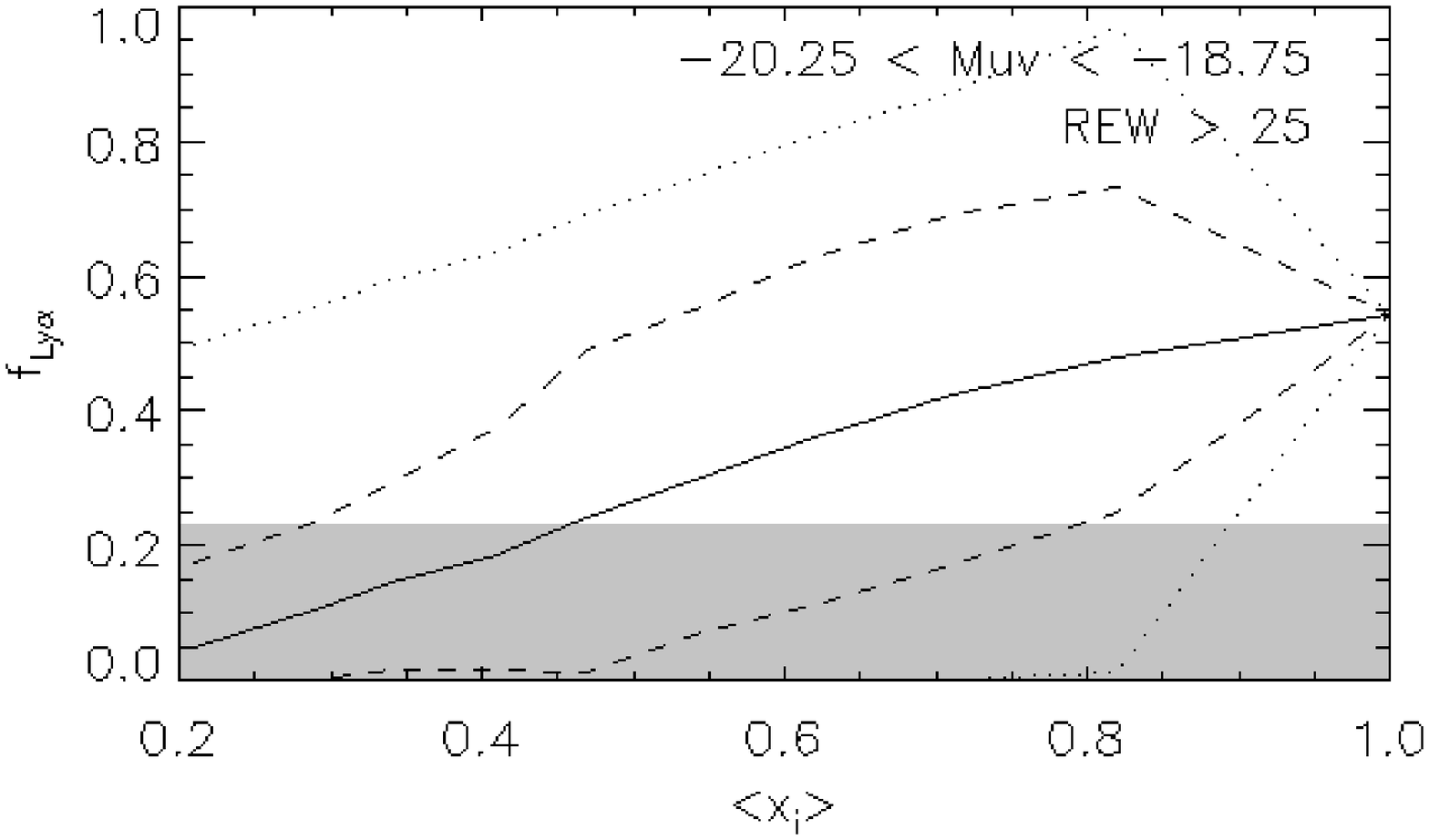}
          \includegraphics[scale=0.30]{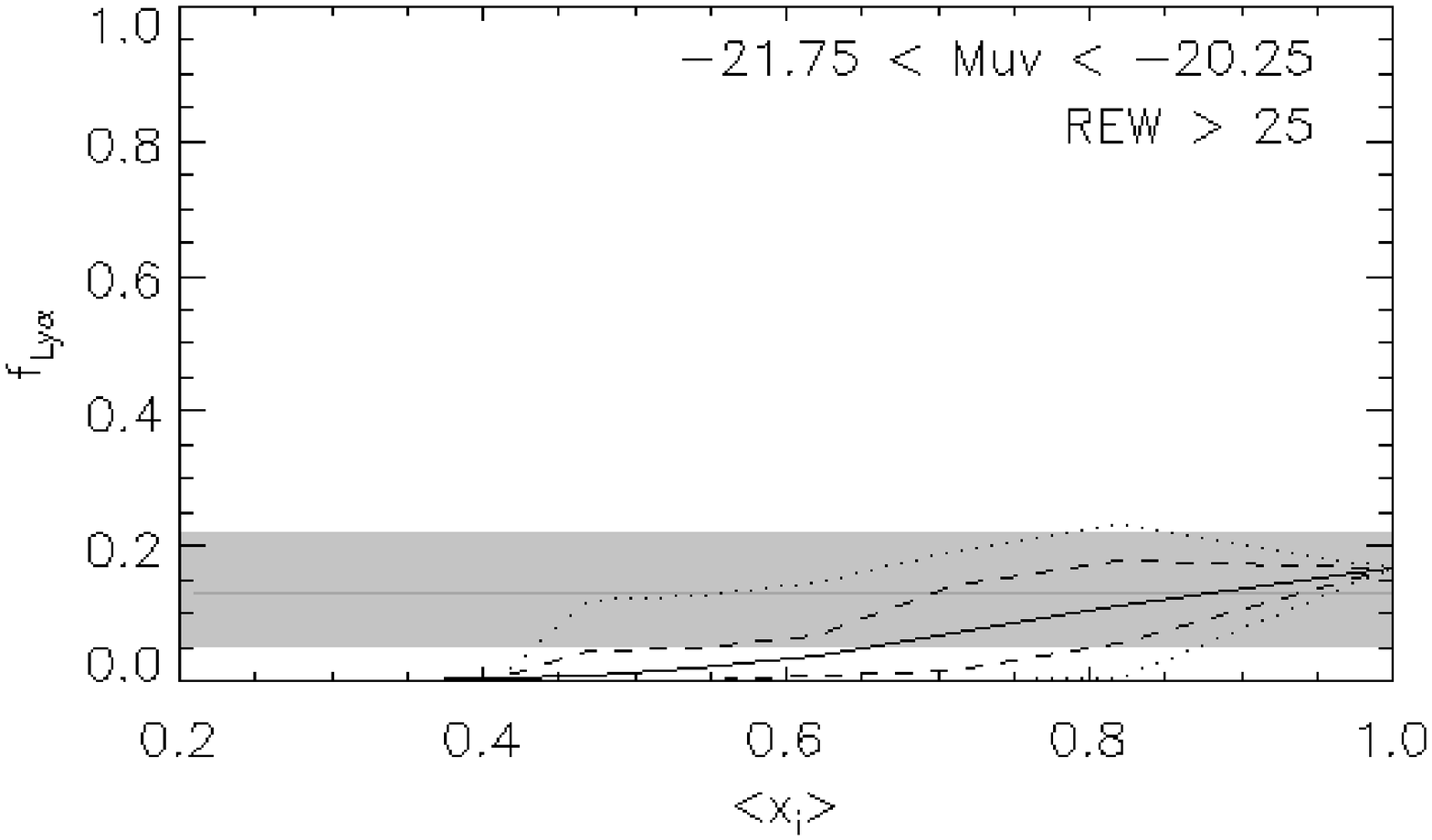}

          \caption{Comparison between mock Ly$\alpha$ fraction ($f_{Ly\alpha}$) measurements and
the observations of \citet{Pentericci11} as a function of the volume-averaged
ionization fraction, $\avg{x_{\rm{i}}}$. This is similar to Fig. \ref{fig:ERS} expect here the shaded bands are
the measurements from \citet{Pentericci11}, with UV faint measurements in the top panel, and UV bright ones in the bottom panel. Likewise,
the simulated observations in this figure mimic those of \citet{Pentericci11}.}

                \label{fig:HAWK}

        \end{center}

\end{figure}

Similarly, we can compare our mock \citet{Pentericci11} observations with their actual measurements. This
is shown in Fig. \ref{fig:HAWK}. Their survey is not as deep
as \citet{Schenker12}, but covers a larger area on the sky. In the
faint UV luminosity bin, they search for Ly$\alpha$ emission around
two LBGs (see Table~\ref{fields}), and find no significant
Ly$\alpha$ emission from either galaxy. This allows them to
place only an upper limit on $f_{Ly\alpha}$ in this luminosity bin.
Their quoted upper limit none the less implies a drop-off towards high redshift in $f_{Ly\alpha}$, when
compared to
the lower redshift measurements of \citet{Stark11}. However, the sample variance in our simulated samples is significant.
In particular, the top 
panel shows that the upper limit on $f_{Ly\alpha}$ translates into a lower limit
on the neutral fraction of $0.2$ at the 68 per cent confidence level.  The spread in $f_{Ly\alpha}$ values is large because only two LBGs are
studied spectroscopically, and there are large variations in the amount of damping wing attenuation suffered by these
two LBGs, depending on whether they reside towards the centre of large ionized regions, at the edge of an ionized region, or in the centre
of a smaller ionized bubble.

For the UV bright galaxies, they do detect some Ly$\alpha$ emission; in this luminosity bin their 
Ly$\alpha$ fraction is lower for small REW ($\rm{REW} > 25 \Ang$) lines than expected from extrapolating
the lower redshift \citet{Stark11} measurements out to $z \sim 7$. However, in our fiducial model the decline in the Ly$\alpha$ fraction
for the UV bright sample is not especially constraining. As discussed previously, these sources tend to have smaller intrinsic
REWs, and so even a small amount of attenuation from the IGM attenuates them out of the observable sample. Indeed the bottom panel of
Fig. \ref{fig:HAWK} shows that the Ly$\alpha$
fraction in the UV bright \citet{Pentericci11} bin is compatible with a fully ionized universe in our fiducial model. As with
the case of \citet{Schenker12}, this bin actually (slightly) disfavours too large a neutral fraction, $\avg{x_{\rm{HI}}} < 0.52$ at the 68 per cent confidence level. Once the neutral fractions is less than $0.4$, $f_{Ly\alpha}$ is consistent with zero; thus, that \citet{Pentericci11} observe any Ly$\alpha$ disfavours a highly neutral universe.

\begin{figure}

        \begin{center}

          \includegraphics[scale=0.3]{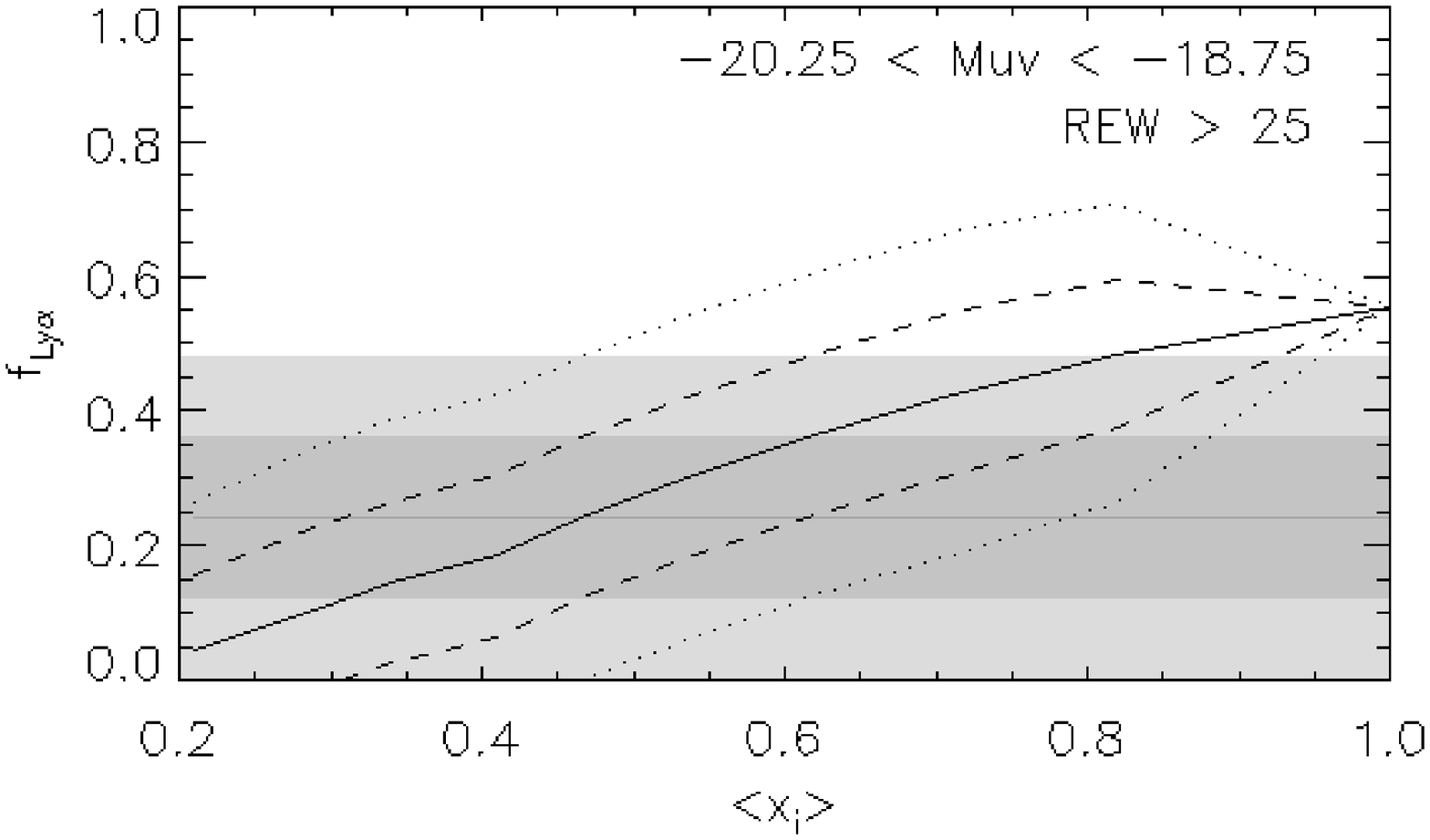}
          \includegraphics[scale=0.3]{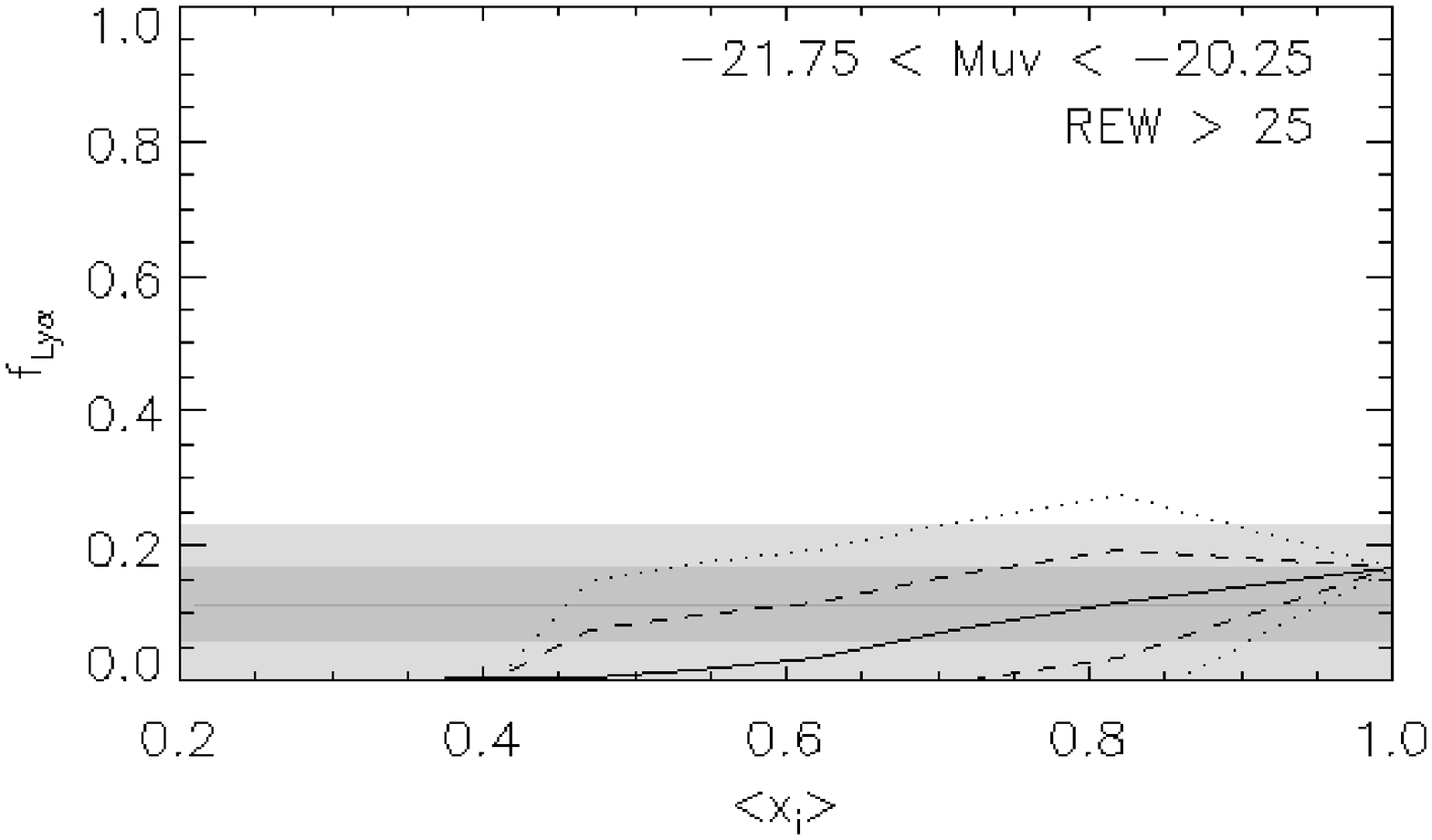}

          \caption{Confidence intervals on $f_{Ly\alpha}$ from combining the measurements of \citet{Schenker12} and \citet{Pentericci11}.
These constraints are compared to our fiducial model,
as a function of $\avg{x_{\rm{i}}}$. Similar to Figs \ref{fig:ERS} and \ref{fig:HAWK}, except here the results reflect the constraints
from combining the separate Ly$\alpha$ fraction measurements. The shaded regions reflect the combined reported errors, while the lines
indicate the simulated average and sample variance contributions to the error budget. The top panel is for the UV faint luminosity bin, while
the bottom panel is for the UV bright bin. }
                \label{fig:combined}

        \end{center}

\end{figure}

Of course,  that both \citet{Pentericci11} and \citet{Schenker12}
observe evidence for a drop in $f_{Ly\alpha}$ near $z \sim 7$
strengthens the case
that these observations probe back into the EoR. In order
to quantify this, we combine the probability distributions for
the separate observations shown in Figs \ref{fig:ERS} and
\ref{fig:HAWK}. The results of this calculation are shown
in Fig \ref{fig:combined}. Here the shaded regions show
the 68 and 95 per cent confidence intervals after combining the
two measurements, assuming that the reported error distributions
obey Gaussian statistics. For the combined UV faint
case, we used directly the combined uncertainty
calculated in \citet{Ono12} which took into account that \citet{Fontana10} was included in the calculations of both \citet{Pentericci11} and \citet{Schenker12}.
As before, the lines show the sample variance
contributions calculated from our simulated models (and not included in the
shaded error budgets).

We do not consider the observations of \citet{Ono12} or \citet{Caruana12} here since we do not expect them to have a significant effect on the combined constraints. While \citet{Ono12} observed a large field, their survey was fairly shallow and focused on a small number of galaxies--only following up on 11 galaxies in the bright bin. Since the faint galaxies place the greatest constraints on the ionized fraction in our fiducial model, a shallow survey, focusing on the bright galaxies, is unlikely to increase the constraints. Further, \citet{Ono12}'s observations are consistent with the increasing trend in $f_{Ly\alpha}$ seen in \citet{Stark11}; \citet{Ono12} do not see a drop in $f_{Ly\alpha}$  at $ z \sim 7$. Together, these characteristics of their observations, mean that not considering their observations does not significantly affect our results. Further, by only observing 11 galaxies, \citet{Ono12} greatly reduce their survey's effective volume. Of course, a deeper survey, with more galaxies sampling the entire large field, would greatly reduce the effects of sample variance and could place significant constraints on the ionized fraction. 

\citet{Caruana12}, on the other hand, focused on an extremely small field of view. The size of their field makes them particularly vulnerable to sample variance. Their observations are consistent with the trend from \citet{Stark11}. Thus, their observations would not place any further constraint on the sample variance.
 
The constraints in the combined case follow the trends found for each separate survey, but
the overall significance is somewhat tightened. As before, the low Ly$\alpha$ fraction
in the UV bright case is consistent with a fully ionized model. The UV faint case does, however,
prefer a partly neutral IGM, but a very large neutral fraction is not required. For example, a model with $\avg{x_i} \sim 0.9$ lies within 
the 95 per cent confidence range.

\subsection{Model dependance}
\label{sec:model_dep}

It is also interesting to explore how the simulated
Ly$\alpha$ fractions depend on the underlying model
for the Ly$\alpha$ emission from LBGs. In particular,
our fiducial model assumes that the intrinsic Ly$\alpha$ REW
is anti correlated with UV luminosity.
A variety of studies seem to support this
relationship. At $z$
\(\simeq\) 3, \citet{Shapley03} report an increase in
the mean REW with fainter luminosity.  \citet{Stark10}
extend this out to higher redshift; for their sample
of LBGs with Ly$\alpha$ emission at \(3 \la z \la
6\), they find that the largest REWs tend to correlate
with fainter UV continuum magnitudes.  These
observations are supported by a host of others
\citep{Ando06, Ouchi08, Vanzella09, Balestra10} that
suggest that among the brightest sources, large
REWs are scarce. Such a correlation has a plausible
explanation; the fainter galaxies may be less dusty than their brighter counterparts,
allowing more of the Ly$\alpha$ emission to be
transmitted. 

While this model does seem to match
observations, there is still not widespread agreement on this point, see, for
instance,
\citet{Kornei10}. 
For LBGs at $ z \sim 3$,  \citet{Kornei10} do not see a significant correlation between continuum luminosity and Ly$\alpha$ emission.
\citet{Nilsson09} argue that a flux
or magnitude-limited survey, like the ones cited
above, will generally observe both fewer instances of strong
Ly$\alpha$ emission and fewer instances of bright LBGs since those are both rare. Thus, in
order to conclude that continuum luminosity and Ly$\alpha$ REWs are anti-correlated one would need very large sample sizes, on the order of 1000 LBGs with measured REWs. In addition,
note that our fiducial model was tuned to match the observations
of \citet{Stark11} in a fully ionized universe. However, there
is still some uncertainty in the post-reionization Ly$\alpha$
fractions, even for UV bright sources.	For example,
\citet{Curtis-Lake12} find a $z \sim 6$ Ly$\alpha$ fraction
that is roughly two times as large as that of \citet{Stark11} for
UV bright sources, and comparable to that of the UV fainter
sources from \citet{Stark11}.

\begin{figure*}

\begin{center} 

	\(
  \begin{array}{ccc}

  \includegraphics[scale=0.28]{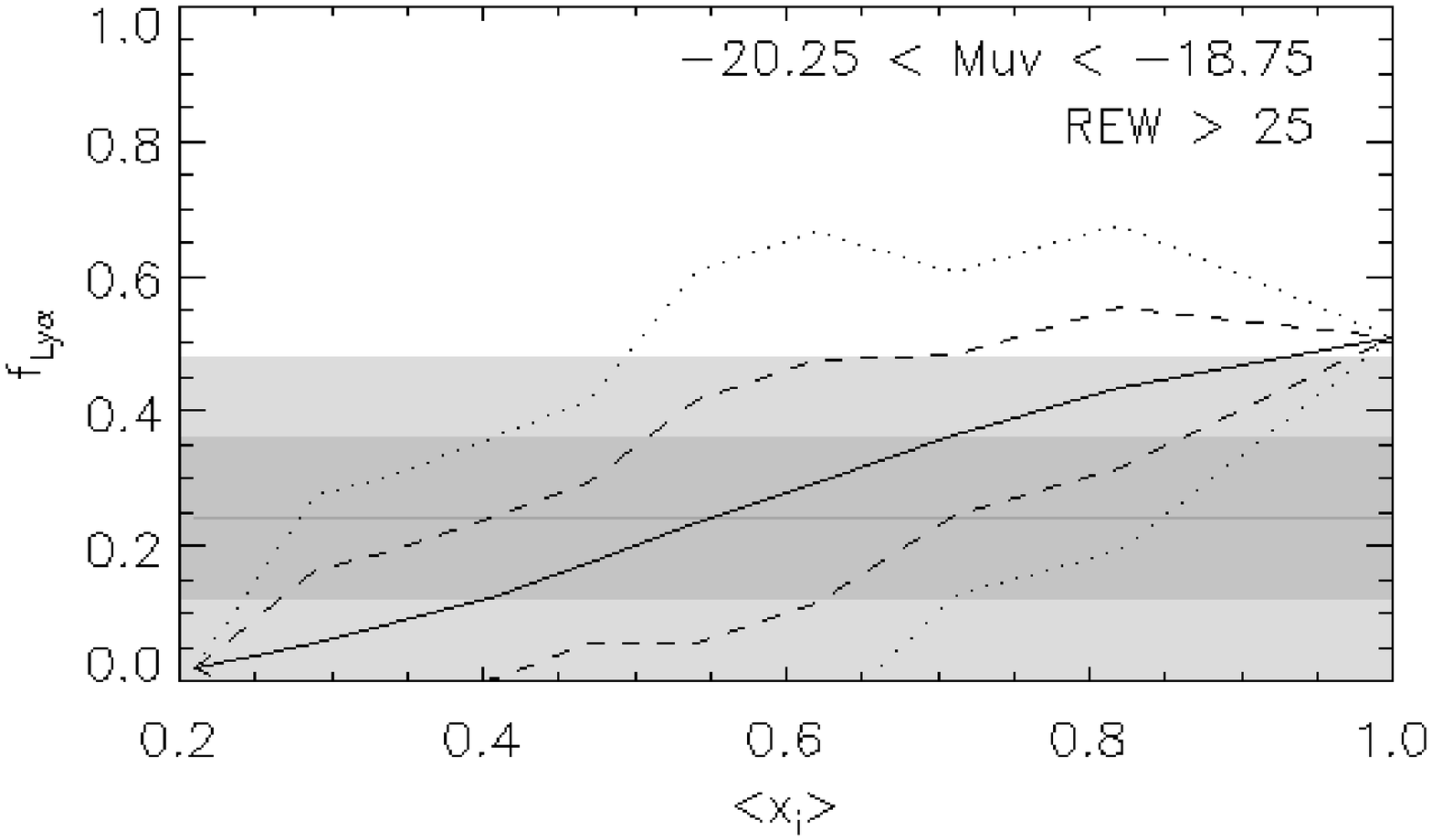} 

  \includegraphics[scale=0.28]{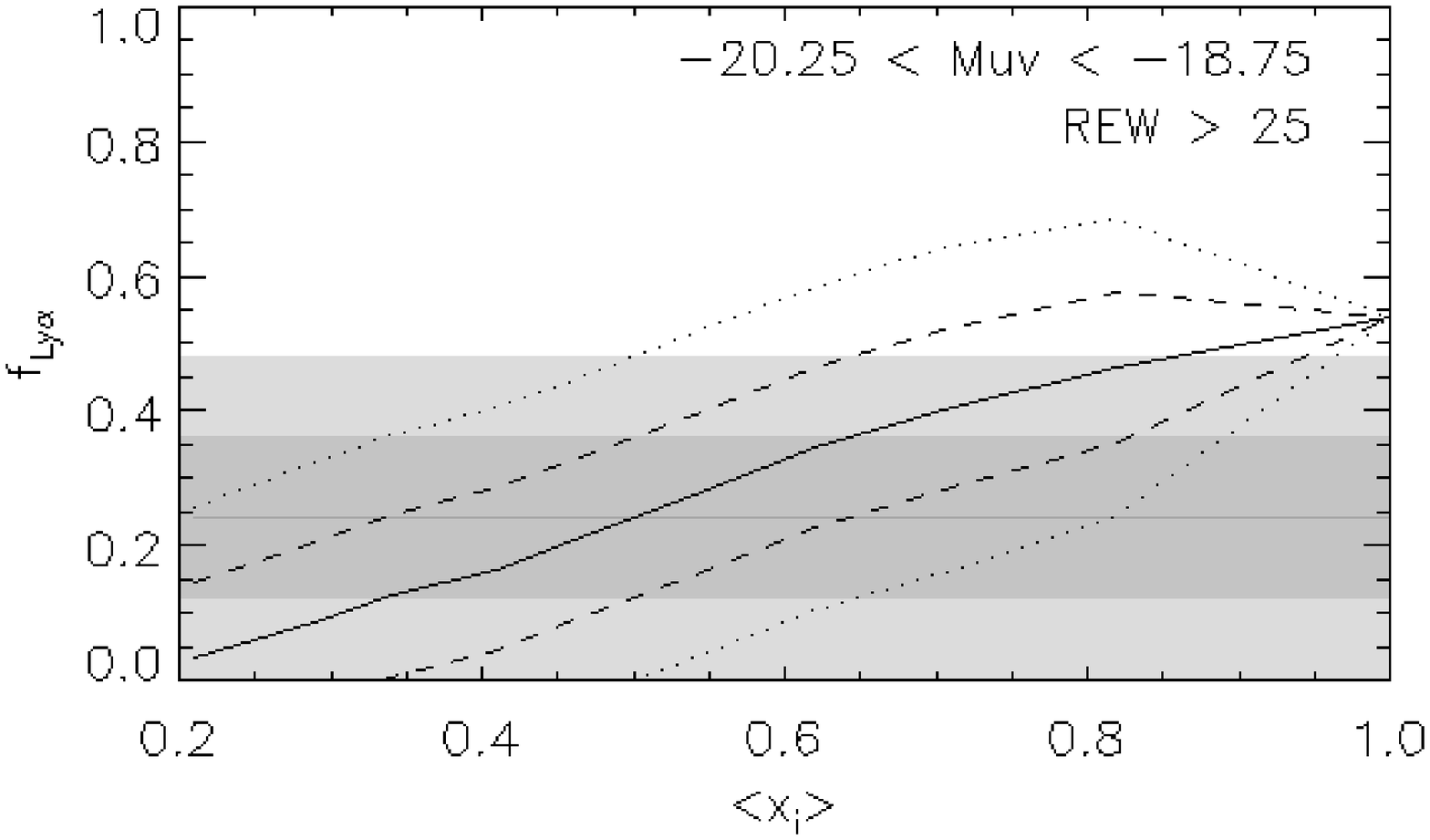}

  \includegraphics[scale=0.28]{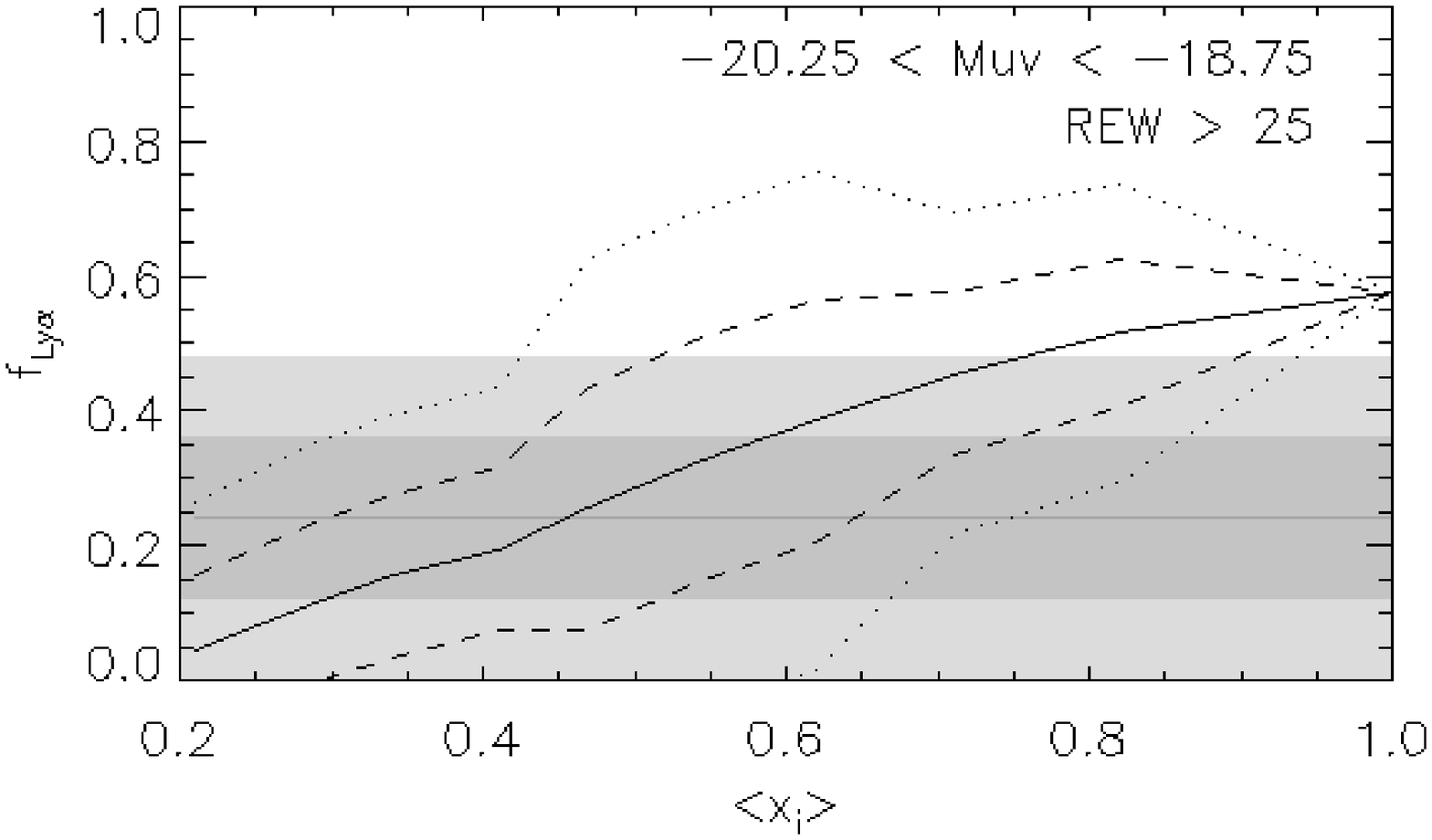} \\

  \includegraphics[scale=0.28]{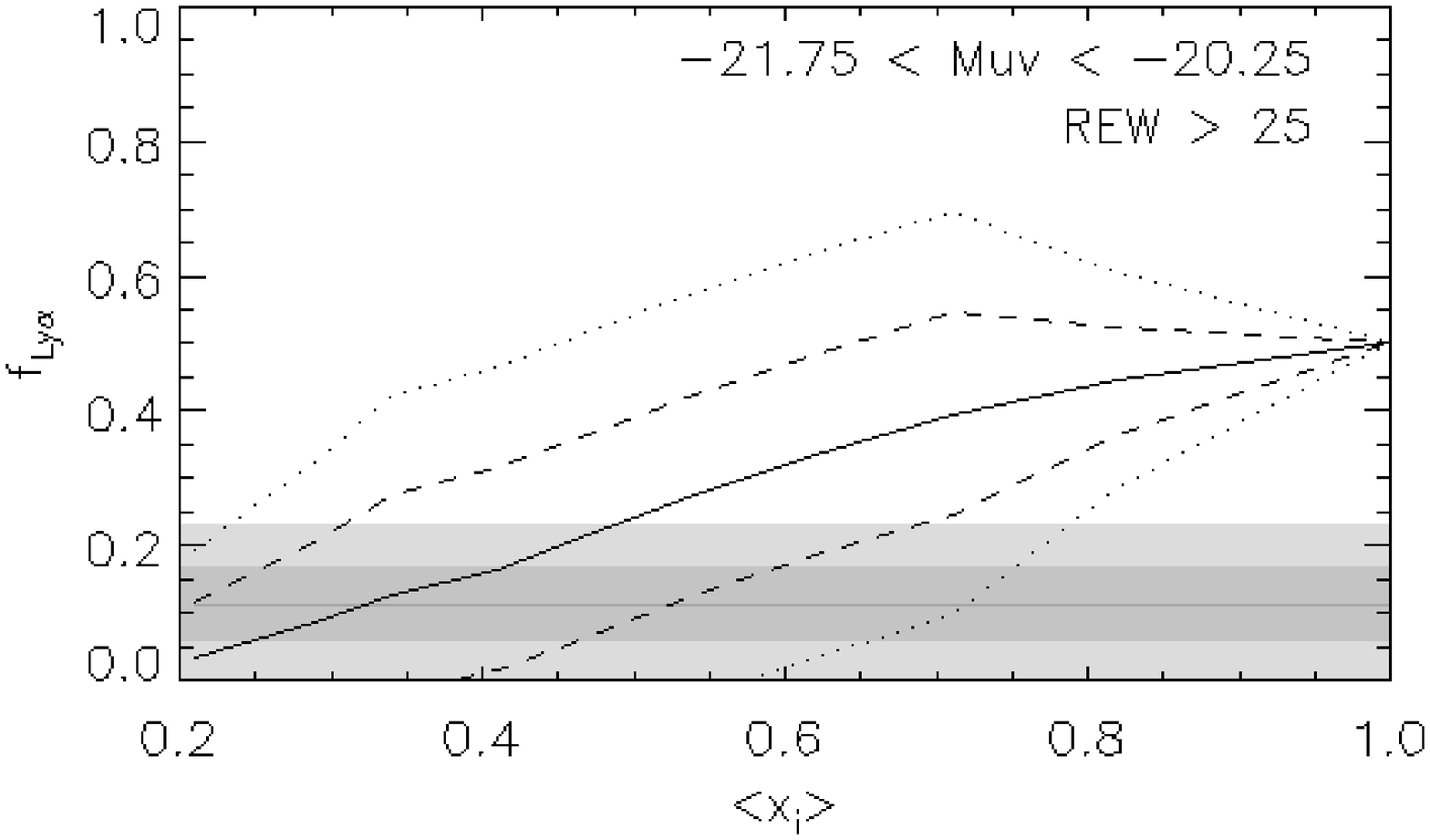}

  \includegraphics[scale=0.28]{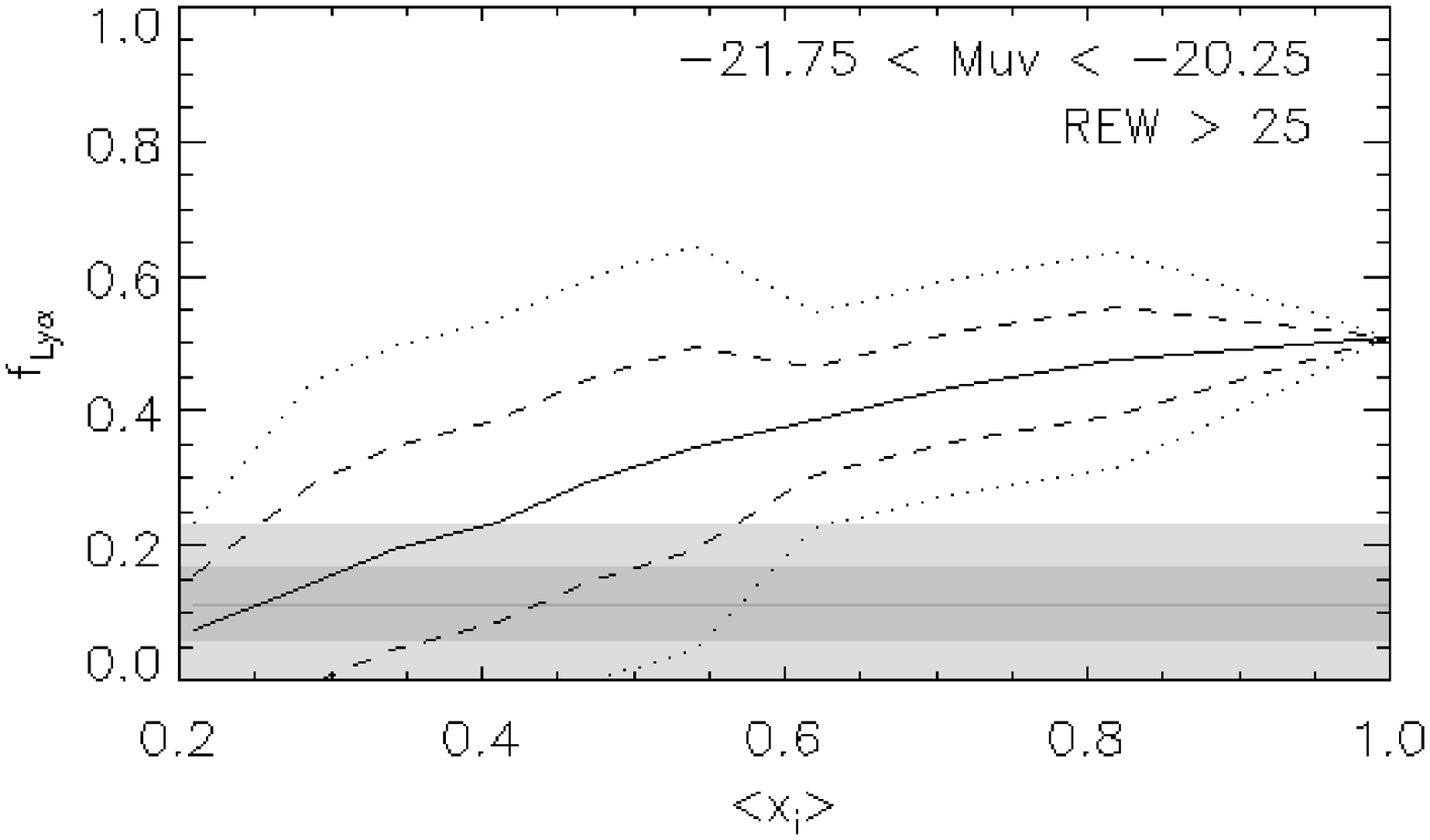}

  \includegraphics[scale=0.28]{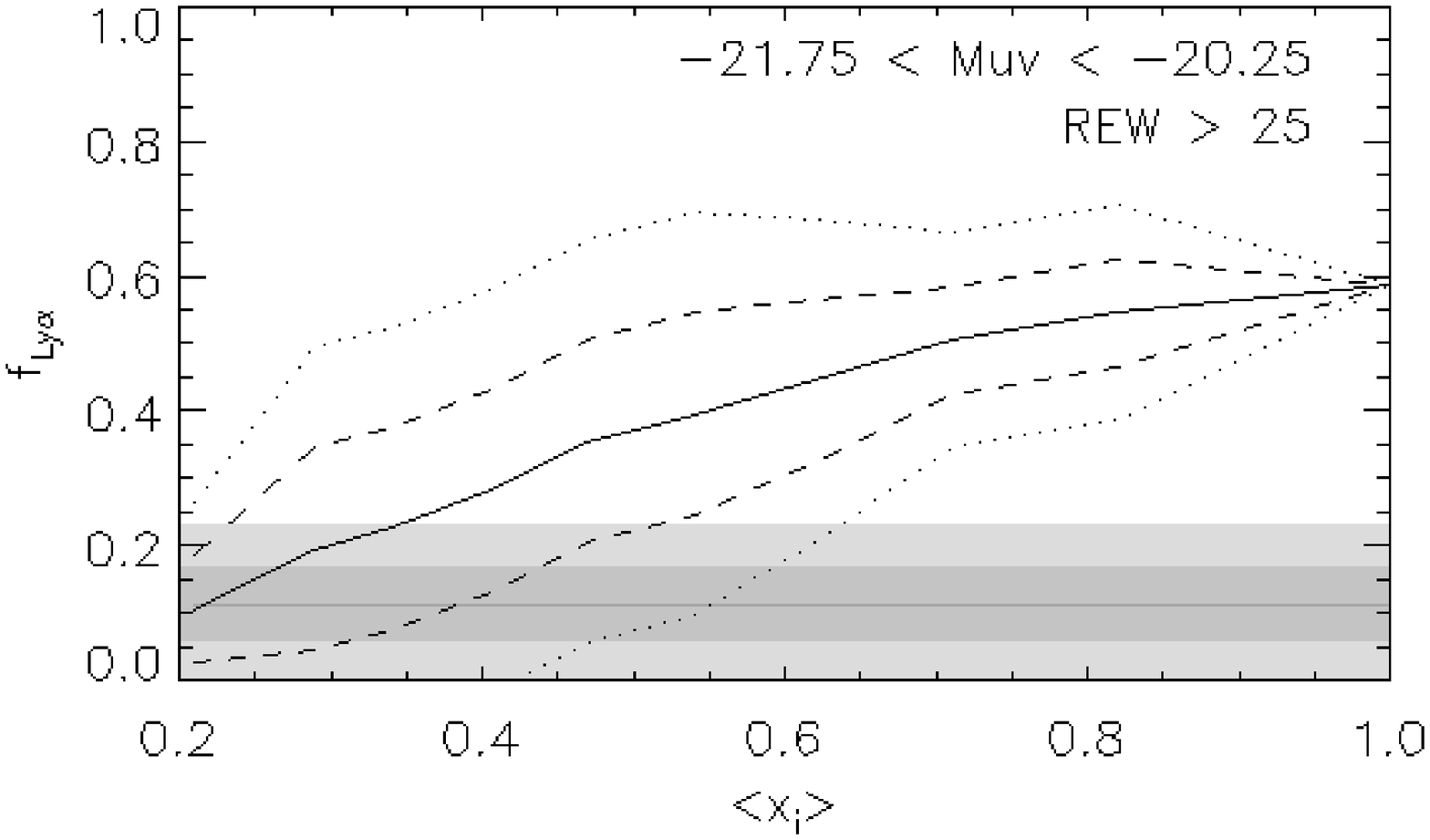}
 
  \end{array} \)
  
  \end{center}

        \caption{Dependence of $f_{Ly\alpha}$ on the LAE model.
Similar to Fig.
          \ref{fig:combined}, but here we consider three
          different models for assigning Ly$\alpha$ REWs to 
          LBGs. Left: there is a weak negative
          correlation between the continuum
          luminosities and the Ly$\alpha$ REWs; the
          fainter galaxies have stronger
          Ly$\alpha$ emission, but the trend is less strong than in our fiducial model. Middle: the
          Ly$\alpha$ REWs are assigned randomly to
          the LBGs, following the distribution
          in \citet{Dijkstra11}.  Right:
          there is a weak positive correlation
          between the continuum luminosities
          and the Ly$\alpha$ REWs; the
          brighter galaxies have stronger
          Ly$\alpha$ emission. The top panels are for UV faint bins, while the
bottom panels show UV bright LBGs.}  

        \label{fig:models}

\end{figure*}

Given the uncertainties in the relationship between LBGs and their
intrinsic Ly$\alpha$ emission, we generate combined constraints along the
lines of Fig. \ref{fig:combined} for three additional LAE models. These additional
models are described in \S \ref{sec:LAE_props} and correspond to cases with:
a weaker negative correlation between intrinsic REW and UV luminosity than in our
fiducial model (with $\rho=-0.11$ from Equation \ref{eq:rho} compared to $\rho=-0.23$ in
our fiducial model); a model where REW and UV luminosity are uncorrelated and a model with a weak positive
correlation ($\rho=0.10$). 

The combined constraints in these additional models are shown in Fig. \ref{fig:models}.
It is reassuring that the results in the UV faint bin appear only weakly dependent on the underlying
model for Ly$\alpha$ emission from LBGs. The UV bright case is, on the other hand, quite sensitive
to the underlying LAE model. In particular, in these models the UV bright Ly$\alpha$ fractions are intrinsically larger
than in our fiducial case, and so significant attenuation is required to explain the lower Ly$\alpha$ fractions
measured in the high UV luminosity bin. However, the same qualitative behaviour is expected in the post-reionization
Universe, and so each of these models is in conflict with the post-reionization ($z \sim 4-6$) measurements of
\citet{Stark11} which do show a significant drop in the Ly$\alpha$ fraction from UV faint to UV bright LBGs.

Our model does not consider the effects of galactic winds on Ly$\alpha$ emission. Backscattering of Ly$\alpha$ emission off of the far side of a galactic outflow
can promote the escape of Ly$\alpha$ photons by shifting them redward of line centre, and strong winds may therefore make Ly$\alpha$ emission more visible even in a highly neutral
universe.
Thus, explaining the drop in $f_{Ly\alpha}$ reported in these surveys would require a more neutral universe than we have argued \citep{Dijkstra11}, strengthening the conclusions of \citet{Schenker12} and \citet{Pentericci11}. However, while strong winds are observed from LBGs at $z \sim 3$ (e.g. \citealt{Shapley03}), the outflow speed
-- and hence the redshift imparted to Ly$\alpha$ photons -- may be substantially smaller around the smaller mass galaxies typical near $z \sim 7$. It is presently
unclear how prevalent strong winds are from $z \sim 7$ LBGs.

Clearly, a better understanding of the relationship
between LBGs and LAEs should help in interpreting
current and future Ly$\alpha$ fraction measurements.
Along these lines, \citet*{Dayal12}
have simulated LBGs and LAEs at $z$ \(\sim\) 6, 7 and 8.
They conclude that LAEs are a diverse subset of
LBGs. While the faintest LBGs do not have Ly\(\alpha\)
emission, LAEs are found distributed throughout all
other categories of LBGs.  Similarly,
\citet{Forero-Romero12} model the relationship between
LBGS and LAEs in the range \(5 \leq z \leq 7\),
concluding that, in order to match the observations of
\citet{Stark11}, the Ly$\alpha$ escape fraction must
be inversely correlated with galaxy
luminosity. Further observational and theoretical work should
help strengthen our understanding of the precise relationship
between these two galaxy populations.

\section{Conclusions}

\begin{figure}

\begin{center}

  \includegraphics[scale=0.41]{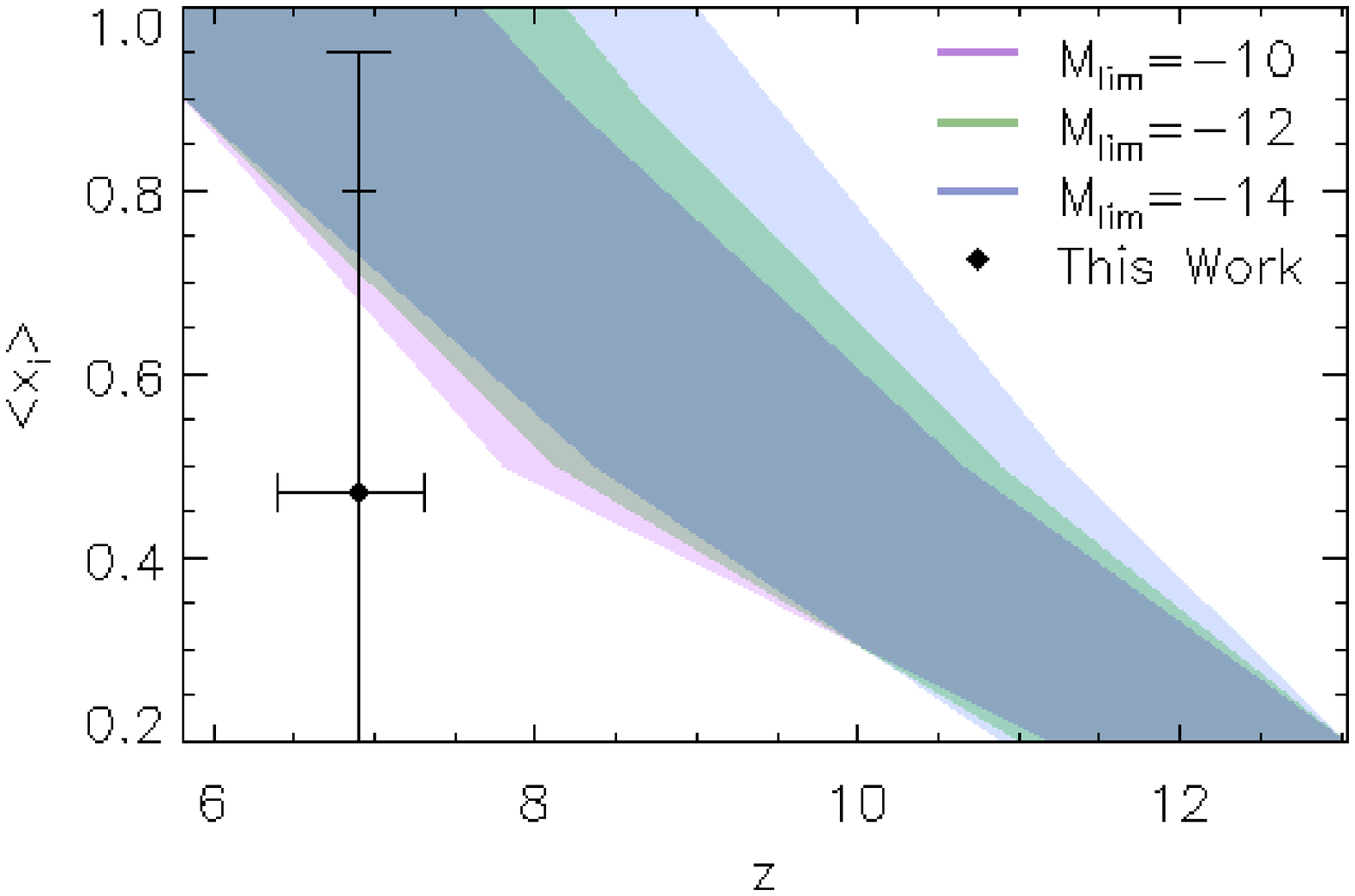}

        \caption{The reionization history inferred from a variety
          of observations. The shaded band shows
          constraints on the ionization fraction as a function
of redshift from \citet{Kuhlen12}, which come from combining
\emph{WMAP}-7 measurements of the optical depth to Thomson scattering, the intensity
of the UV background inferred from the $z \sim$ 2-5 Ly$\alpha$ forest, and 
measurements of the UV galaxy luminosity function at high redshift. The luminosity functions are
extrapolated down to faint luminosities; the different shaded bands show the dependence on
the limiting UV magnitude ($M_{\rm{lim}}$) out to which this extrapolation is carried out.
The point with error bars shows the constraint we infer from the Ly$\alpha$ fraction measurements of \citet{Schenker12}
and \citet{Pentericci11}. The lower vertical error bar shows the 68 per cent confidence interval, while the
upper vertical error bar shows the 95 per cent confidence interval. The horizontal error bar gives the redshift uncertainty
for the LBGs used in these measurements.}

        \label{fig:conclusion}

\end{center}

\end{figure}

\begin{figure}

\begin{center}

  \includegraphics[scale=0.3]{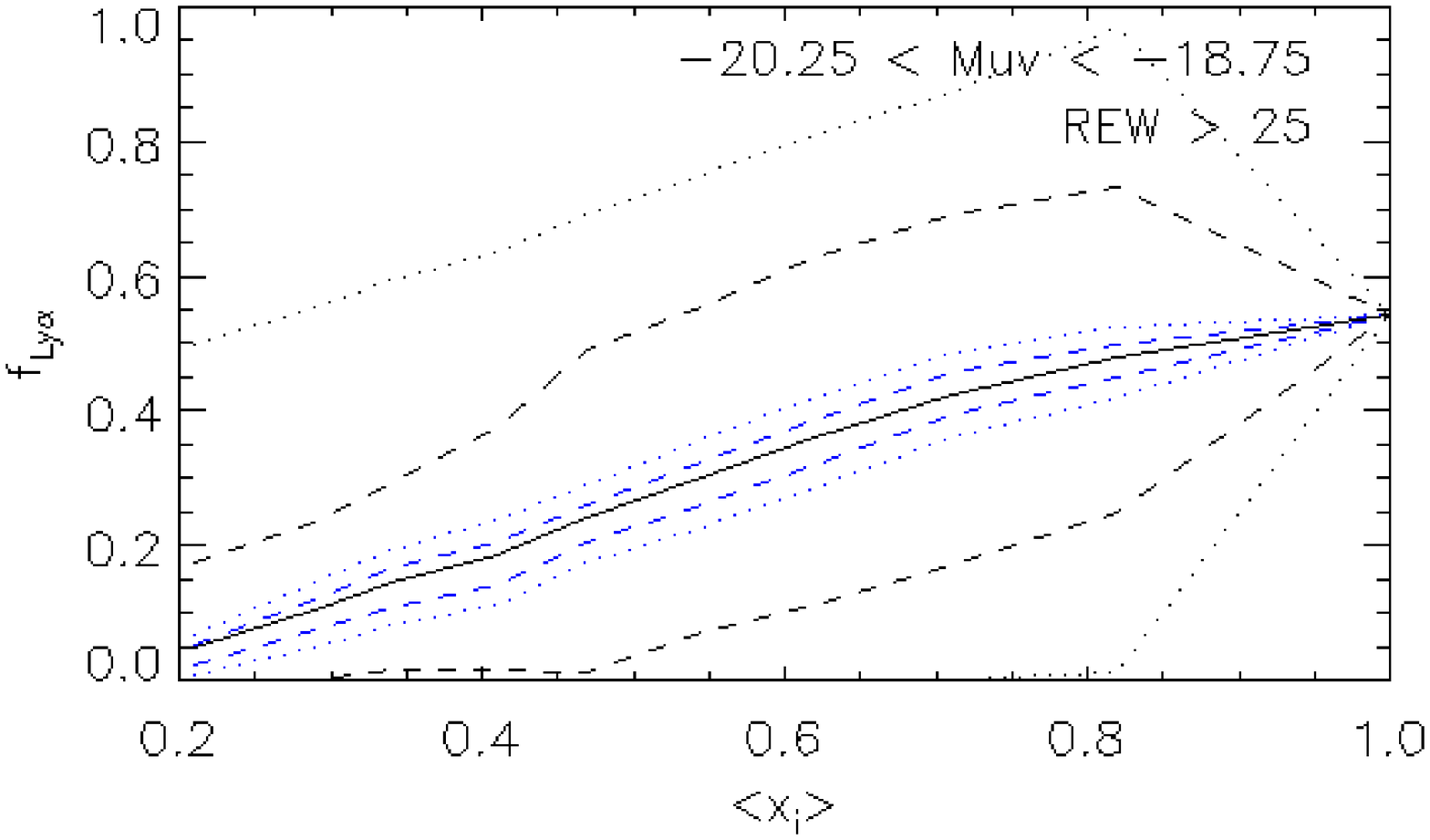}
  
  	\caption{The impact of improved sampling of the \citet{Pentericci11} fields on the sample variance error.
\citet{Pentericci11} looked for Ly$\alpha$ emission from two faint LBGs. This leads to the large sample variance, shown in the black dashed and dotted lines, at 68 and 95 per cent, respectively (identical to the lines in the top panel of Fig. \ref{fig:HAWK}). The blue dashed and dotted lines show that the sample variance would dramatically shrink if a survey was done to search for Ly$\alpha$ emission from {\em all} of the LBGs in their fields.}
	
	\label{fig:future}
	
\end{center}

\end{figure}

In order to summarize our findings and put them in the larger context of
our current understanding of the EoR, we use the results of Fig. \ref{fig:combined}
to compute the likelihood function for $\avg{x_{\rm{i}}}(z \sim 7)$ given the results
of both \citet{Schenker12} and \citet{Pentericci11}.  Fig. \ref{fig:conclusion} shows
the resulting 68 and 95 per cent confidence regions for $\avg{x_{\rm{i}}}$, as compared
to constraints from \cite{Kuhlen12}, which combine recent \emph{WMAP}-7 measurements of
the Thomson scattering optical depth, along with observations of the $z \sim 2-5$ Ly$\alpha$ forest,
and UV galaxy luminosity function measurements.  Although the central value of $\avg{x_{\rm{i}}}(z \sim 7)$ we infer from the
$f_{Ly\alpha}$ observations is
outside of the range implied by the other measurements, accounting for sample variance and other
sources of uncertainty in the $f_{Ly\alpha}$ measurements, we find that these various observations are in fact in broad agreement
with each other. Interestingly, we conclude that a fully ionized IGM is disfavoured by the $f_{Ly\alpha}$ measurements, 
although a highly neutral IGM ($\avg{x_{\rm{i}}} \leq \sim 0.5$) is not required. Taken together with spectroscopic observations 
of a $z=7.1$ quasar from \citet{Mortlock:2011va}, the case that existing measurements probe into the EoR is strengthening.

It is also interesting to note that recent and improved measurements of the UV galaxy luminosity function from the Ultra Deep Field (UDF) survey  and Cosmic Assembly Near-Infrared Deep Extragalactic Survey (CANDELS) favour somewhat later reionization than in the \citet{Kuhlen12} analysis \citep{Oesch13, Robertson13}.
For instance,  extrapolating the measured luminosity functions down to $M_{\rm UV} < -13$, \citet{Robertson13}
favours $z \sim 7.5$ for the redshift at which the Universe is $50$ per cent neutral by volume (see their Fig. 5).
These results are more easily harmonized with the preferred (central) value of $\avg{x_{\rm{i}}}(z \sim 7)$ inferred from the $f_{Ly\alpha}$ observations.

We also find that the constraints from the Ly$\alpha$ fraction measurements depend somewhat on the model relating the properties of the LBGs and LAEs. In our fiducial model (shown here in Fig. \ref{fig:conclusion}), the intrinsic REW of Ly$\alpha$ emission from LBGs is anti-correlated with UV luminosity. In models without this anti correlation, the lower Ly$\alpha$ fractions among UV bright galaxies at both $z \sim 7$ and at lower redshifts ($z \sim 4-6$) are hard to understand.

\citet{Bolton13} recently pointed out another effect that
may also help reconcile the Ly$\alpha$ fraction measurements
with other constraints on reionization. These authors suggest
that the observed drops in the Ly$\alpha$ fraction towards
high redshift may be driven mostly by strong evolution in the
number of dense, optically thick absorbers in the vicinity
of the observed LBGs, rather than reflecting prominent changes
in the ionization state of the diffuse IGM. This effect would not
be well captured in the simulations used in our analysis, since
these dense absorbers are notoriously difficult to resolve in large volume
reionization simulations. This effect would also, however, presumably show
large field-to-field variations, similar to the spatial variations
in the attenuation from the diffuse gas studied here. More detailed
models will likely be necessary to disentangle the relative impact
of optically thick, circumgalactic absorbers and the more diffuse IGM
considered here.

More importantly, the prospects for improved observations are very good, and especially
tantalizing given the current hints that $z \sim 7$ observations may
probe into the EoR. Larger and deeper surveys
for LBGs and LAEs should help clarify the interpretation of current observations. In particular, as shown in Fig. \ref{fig:future}, observing down to $M_{UV} = -18.75 $ over an area as large as that observed by \citet{Pentericci11} would be enough to greatly reduce the effects of sample variance. Fully sampling such an area, however, would mean spectroscopically observing on the order of 100s of LBGs, looking for Ly$\alpha$ emission. However, even just sampling five galaxies per field (20 galaxies total) is enough to `fill in' the sparse sampling and reduce the sample variance by a factor of 2. Alternatively, the use of narrow-band filters to select the LBGs that are also LAEs should make observing 100s feasible. 

The spatial fluctuations in the Ly$\alpha$ fraction studied here are presently a nuisance;  however,
they should {\em ultimately provide a very interesting and distinctive signature of reionization},
provided they can be mapped out over large volumes of the Universe. Mapping these fluctuations will require both larger fields of view and more fully sampled fields than currently available. This should, nonetheless, be possible using upcoming observations from the Hyper Suprime-Cam on the
Subaru telescope, which will observe LAEs out to $z =7.3$ over more than $5$  deg$^2$.

\section{Acknowledgements}

We thank Matt McQuinn for providing the reionization simulations used in our analysis and for comments on a draft.
JT and AL were supported by NASA grant NNX12AC97G.

\bibliographystyle{astron}
\bibliography{reionization_8_10_13}

\begin{thebibliography}{}

\bibitem[\protect\astroncite{Ade et~al.}{2013}]{Ade:2013zuv}
Ade, P. et~al.: 2013,
\newblock {\em ArXiv e-prints}

\bibitem[\protect\astroncite{Ando et~al.}{2006}]{Ando06}
Ando, M., Ohta, K., Iwata, I., et~al.: 2006,
\newblock {\em {ApJL}} {\bf 645}, L9

\bibitem[\protect\astroncite{Arnouts et~al.}{1999}]{Arnouts99}
Arnouts, S., D'Odorico, S.~Cristiani, S., Zaggia, S., Fontana, A., and
  Giallongo, E.: 1999,
\newblock {\em {A\&A}} {\bf 341}, 641

\bibitem[\protect\astroncite{Balestra et~al.}{2010}]{Balestra10}
Balestra, I., Mainieri, V., Popesso, P., et~al.: 2010,
\newblock {\em A\&A} {\bf 512}, A12

\bibitem[\protect\astroncite{Barkana and Loeb}{2001}]{Barkana01}
Barkana, R. and Loeb, A.: 2001,
\newblock {\em {P}hys. {R}ep.} {\bf 349}, 125

\bibitem[\protect\astroncite{{Becker} and {Bolton}}{2013}]{Becker13}
{Becker}, G.~D. and {Bolton}, J.~S.: 2013,
\newblock {\em ArXiv e-prints}

\bibitem[\protect\astroncite{{Blanc} et~al.}{2011}]{Blanc11}
{Blanc}, G.~A., {Adams}, J.~J., {Gebhardt}, K., et~al.: 2011,
\newblock {\em ApJ} {\bf 736}, 31

\bibitem[\protect\astroncite{Bolton and Haehnelt}{2013}]{Bolton13}
Bolton, J.~S. and Haehnelt, M.: 2013,
\newblock {\em {MNRAS}} {\bf 429}, 1695

\bibitem[\protect\astroncite{Bolton and Haehnelt}{2007}]{Bolton:2007fw}
Bolton, J.~S. and Haehnelt, M.~G.: 2007,
\newblock {\em MNRAS}

\bibitem[\protect\astroncite{{Bouwens} et~al.}{2011}]{Bouwens11}
{Bouwens}, R.~J., {Illingworth}, G.~D., {Oesch}, P.~A., et~al.: 2011,
\newblock {\em ApJ} {\bf 737}, 90

\bibitem[\protect\astroncite{{Caruana} et~al.}{2012}]{Caruana12}
{Caruana}, J., {Bunker}, A.~J., {Wilkins}, S.~M., et~al.: 2012,
\newblock {\em MNRAS} {\bf 427}, 3055

\bibitem[\protect\astroncite{{Castellano} et~al.}{2010}]{Castellano10}
{Castellano}, M., {Fontana}, A., {Paris}, D., et~al.: 2010,
\newblock {\em A\&A} {\bf 524}, A28

\bibitem[\protect\astroncite{{Cl{\'e}ment} et~al.}{2012}]{Clement12}
{Cl{\'e}ment}, B., {Cuby}, J.-G., {Courbin}, F., et~al.: 2012,
\newblock {\em A\&A} {\bf 538}, A66

\bibitem[\protect\astroncite{{Curtis-Lake} et~al.}{2012}]{Curtis-Lake12}
{Curtis-Lake}, E., {McLure}, R.~J., {Pearce}, H.~J., et~al.: 2012,
\newblock {\em MNRAS} {\bf 422}, 1425

\bibitem[\protect\astroncite{{Dayal} and {Ferrara}}{2012}]{Dayal12}
{Dayal}, P. and {Ferrara}, A.: 2012,
\newblock {\em MNRAS} {\bf 421}, 2568

\bibitem[\protect\astroncite{Dijkstra et~al.}{2007}]{Dijkstra:2007nj}
Dijkstra, M., Lidz, A., and Wyithe, S.: 2007,
\newblock {\em MNRAS} {\bf 377}, 1175

\bibitem[\protect\astroncite{Dijkstra et~al.}{2011}]{Dijkstra11}
Dijkstra, M., Mesinger, A., and Wyithe, J. S.~B.: 2011,
\newblock {\em {MNRAS}} {\bf 414}, 2139

\bibitem[\protect\astroncite{Dijkstra et~al.}{2006}]{Dijkstra:2006xx}
Dijkstra, M., Wyithe, S., and Haiman, Z.: 2006,
\newblock {\em MNRAS}

\bibitem[\protect\astroncite{{Duval} et~al.}{2013}]{Duval13}
{Duval}, F., {Schaerer}, D., {{\"O}stlin}, G., and {Laursen}, P.: 2013,
\newblock {\em ArXiv e-prints}

\bibitem[\protect\astroncite{Fan et~al.}{2006}]{Fan:2005es}
Fan, X.-H., Strauss, M.~A., Becker, R.~H., et~al.: 2006,
\newblock {\em AJ} {\bf 132}, 117

\bibitem[\protect\astroncite{Fontana et~al.}{2000}]{Fontana00}
Fontana, A., D'Odorico, S., Poli, F., et~al.: 2000,
\newblock {\em {AJ}} {\bf 120}, 2206

\bibitem[\protect\astroncite{Fontana et~al.}{2010}]{Fontana10}
Fontana, A., Vanzella, E.~Pentericci, L., Castellano, M., et~al.: 2010,
\newblock {\em {ApJL}} {\bf 725}, 205

\bibitem[\protect\astroncite{{Forero-Romero} et~al.}{2012}]{Forero-Romero12}
{Forero-Romero}, J.~E., {Yepes}, G., {Gottl{\"o}ber}, S., and {Prada}, F.:
  2012,
\newblock {\em MNRAS} {\bf 419}, 952

\bibitem[\protect\astroncite{Furlanetto et~al.}{2006}]{Furlanetto:2005ir}
Furlanetto, S.~R., Zaldarriaga, M., and Hernquist, L.: 2006,
\newblock {\em MNRAS} {\bf 365}, 1012

\bibitem[\protect\astroncite{Giavalisco et~al.}{2004}]{Giavalisco04}
Giavalisco, M., Ferguson, H.~C., Koekemoer, A.~M., et~al.: 2004,
\newblock {\em {ApJL}} {\bf 600}, 93

\bibitem[\protect\astroncite{{Gronwall} et~al.}{2007}]{Gronwall07}
{Gronwall}, C., {Ciardullo}, R., {Hickey}, T., et~al.: 2007,
\newblock {\em ApJ} {\bf 667}, 79

\bibitem[\protect\astroncite{Hathi et~al.}{2010}]{Hathi10}
Hathi, N.~P., Ryan, R.~E., Cohen, S.~H., et~al.: 2010,
\newblock {\em {ApJ}} {\bf 720}, 1708

\bibitem[\protect\astroncite{{Hibon} et~al.}{2010}]{Hibon10}
{Hibon}, P., {Cuby}, J.-G., {Willis}, J., et~al.: 2010,
\newblock {\em A\&A} {\bf 515}, A97

\bibitem[\protect\astroncite{{Hibon} et~al.}{2012}]{Hibon12}
{Hibon}, P., {Kashikawa}, N., {Willott}, C., {Iye}, M., and {Shibuya}, T.:
  2012,
\newblock {\em ApJ} {\bf 744}, 89

\bibitem[\protect\astroncite{{Hibon} et~al.}{2011}]{Hibon11}
{Hibon}, P., {Malhotra}, S., {Rhoads}, J., and {Willott}, C.: 2011,
\newblock {\em ApJ} {\bf 741}, 101

\bibitem[\protect\astroncite{Hu et~al.}{2010}]{Hu:2010um}
Hu, E., Cowie, L., Barger, A., et~al.: 2010,
\newblock {\em ApJ} {\bf 725}, 394

\bibitem[\protect\astroncite{{Jensen} et~al.}{2013}]{Jensen:2012uk}
{Jensen}, H., {Laursen}, P., {Mellema}, G., et~al.: 2013,
\newblock {\em MNRAS} {\bf 428}, 1366

\bibitem[\protect\astroncite{Kashikawa et~al.}{2006}]{Kashikawa:2006pb}
Kashikawa, N., Shimasaku, K., Malkan, M.~A., et~al.: 2006,
\newblock {\em ApJ} {\bf 648}, 7

\bibitem[\protect\astroncite{Kashikawa et~al.}{2011}]{Kashikawa:2011dz}
Kashikawa, N., Shimasaku, K., Matsuda, Y., et~al.: 2011,
\newblock {\em ApJ} {\bf 734}, 119

\bibitem[\protect\astroncite{Kornei et~al.}{2010}]{Kornei10}
Kornei, K.~A., Shapley, A.~E., Erb, D.~K., et~al.: 2010,
\newblock {\em {ApJ}} {\bf 711}, 693

\bibitem[\protect\astroncite{{Krug} et~al.}{2012}]{Krug12}
{Krug}, H.~B., {Veilleux}, S., {Tilvi}, V., et~al.: 2012,
\newblock {\em ApJ} {\bf 745}, 122

\bibitem[\protect\astroncite{{Kuhlen} and
  {Faucher-Gigu{\`e}re}}{2012}]{Kuhlen12}
{Kuhlen}, M. and {Faucher-Gigu{\`e}re}, C.-A.: 2012,
\newblock {\em MNRAS} {\bf 423}, 862

\bibitem[\protect\astroncite{Larson et~al.}{2011}]{Larson:2010gs}
Larson, D., Dunkley, J., Hinshaw, G., et~al.: 2011,
\newblock {\em ApJS} {\bf 192}, 16

\bibitem[\protect\astroncite{Lehnert and Bremer}{2003}]{Lehnert03}
Lehnert, M.~D. and Bremer, M.: 2003,
\newblock {\em ApJ} {\bf 493}, 630

\bibitem[\protect\astroncite{Lidz et~al.}{2007}]{Lidz:2007mz}
Lidz, A., McQuinn, M., and Zaldarriaga, M.: 2007,
\newblock {\em ApJ} {\bf 670}, 39

\bibitem[\protect\astroncite{Madau et~al.}{1998}]{Madau98}
Madau, P., Pozzetti, L., and Dickinson, M.: 1998,
\newblock {\em {ApJ}} {\bf 498}, 106

\bibitem[\protect\astroncite{{McGreer} et~al.}{2011}]{McGreer:2011dm}
{McGreer}, I.~D., {Mesinger}, A., and {Fan}, X.: 2011,
\newblock {\em MNRAS} {\bf 415}, 3237

\bibitem[\protect\astroncite{McLure et~al.}{2011}]{McLure11}
McLure, R.~J., Dunlop, J.~S., de~Ravel, L., et~al.: 2011,
\newblock {\em {MNRAS}} {\bf 418}, 2074

\bibitem[\protect\astroncite{McQuinn et~al.}{2007a}]{McQuinn07b}
McQuinn, M., Hernquist, L., Zaldarriaga, M., and Dutta, S.: 2007a,
\newblock {\em MNRAS} {\bf 381}, 75

\bibitem[\protect\astroncite{McQuinn et~al.}{2007b}]{McQuinn07a}
McQuinn, M., Lidz, A., Zahn, O., et~al.: 2007b,
\newblock {\em {MNRAS}} {\bf 377}, 1043

\bibitem[\protect\astroncite{{Mesinger}}{2010}]{Mesinger:2010mv}
{Mesinger}, A.: 2010,
\newblock {\em MNRAS} {\bf 407}, 1328

\bibitem[\protect\astroncite{{Mesinger} and
  {Furlanetto}}{2008}]{Mesinger:2008jr}
{Mesinger}, A. and {Furlanetto}, S.~R.: 2008,
\newblock {\em MNRAS} {\bf 386}, 1990

\bibitem[\protect\astroncite{Miralda-Escude}{1998}]{MiraldaEscude:1997qb}
Miralda-Escude, J.: 1998,
\newblock {\em ApJ} {\bf 501}, 15

\bibitem[\protect\astroncite{Miralda-Escude}{2003}]{MiraldaEscude:2002yd}
Miralda-Escude, J.: 2003,
\newblock {\em ApJ} {\bf 597}, 66

\bibitem[\protect\astroncite{Mortlock et~al.}{2011}]{Mortlock:2011va}
Mortlock, D.~J., Warren, S.~J., Venemans, B.~P., et~al.: 2011,
\newblock {\em Nature} {\bf 474}, 616

\bibitem[\protect\astroncite{Nilsson et~al.}{2009}]{Nilsson09}
Nilsson, K.~K., M\"{o}ller-Nilsson, O., M{\o}ller, P., Fynbo, J. P.~U., and
  Shapley, A.~E.: 2009,
\newblock {\em {MNRAS}} {\bf 400}, 232

\bibitem[\protect\astroncite{{Oesch} et~al.}{2013}]{Oesch13}
{Oesch}, P.~A., {Bouwens}, R.~J., {Illingworth}, G.~D., et~al.: 2013,
\newblock {\em ApJ} {\bf 773}, 75

\bibitem[\protect\astroncite{Ono et~al.}{2012}]{Ono12}
Ono, Y., Ouchi, M., Mobasher, B., et~al.: 2012,
\newblock {\em {ApJ}} 744

\bibitem[\protect\astroncite{Ouchi et~al.}{2009}]{Ouchi09}
Ouchi, M., Mobasher, B., Shimasaku, K., et~al.: 2009,
\newblock {\em ApJ} {\bf 706}, 1136

\bibitem[\protect\astroncite{Ouchi et~al.}{2008}]{Ouchi08}
Ouchi, M., Shimasaku, K., Akiyama, M., et~al.: 2008,
\newblock {\em {ApJS}} {\bf 176}, 301

\bibitem[\protect\astroncite{{Ouchi} et~al.}{2010}]{Ouchi10}
{Ouchi}, M., {Shimasaku}, K., {Furusawa}, H., et~al.: 2010,
\newblock {\em ApJ} {\bf 723}, 869

\bibitem[\protect\astroncite{{Overzier} et~al.}{2006}]{Overzier06}
{Overzier}, R.~A., {Bouwens}, R.~J., {Illingworth}, G.~D., and {Franx}, M.:
  2006,
\newblock {\em ApJL} {\bf 648}, L5

\bibitem[\protect\astroncite{Pentericci et~al.}{2011}]{Pentericci11}
Pentericci, L., Fontana, A., Vanzella, E., et~al.: 2011,
\newblock {\em {ApJ}} 743

\bibitem[\protect\astroncite{{Pritchard} et~al.}{2010}]{Pritchard:2010nm}
{Pritchard}, J.~R., {Loeb}, A., and {Wyithe}, J.~S.~B.: 2010,
\newblock {\em MNRAS} {\bf 408}, 57

\bibitem[\protect\astroncite{{Robertson} et~al.}{2013}]{Robertson13}
{Robertson}, B.~E., {Furlanetto}, S.~R., {Schneider}, E., et~al.: 2013,
\newblock {\em ApJ} {\bf 768}, 71

\bibitem[\protect\astroncite{Santos}{2004}]{Santos:2003pc}
Santos, M.~R.: 2004,
\newblock {\em MNRAS} {\bf 349}, 1137

\bibitem[\protect\astroncite{Schenker et~al.}{2012}]{Schenker12}
Schenker, M.~A., Stark, D.~P., Ellis, R.~S., et~al.: 2012,
\newblock {\em {ApJ}} 744

\bibitem[\protect\astroncite{Shapley et~al.}{2003}]{Shapley03}
Shapley, A.~E., C., S.~C., Pettini, M., and Adelberger, K.~L.: 2003,
\newblock {\em {ApJ}} {\bf 588}, 65

\bibitem[\protect\astroncite{Springel}{2005}]{Springel05}
Springel, V.: 2005,
\newblock {\em {MNRAS}} {\bf 364}, 1105

\bibitem[\protect\astroncite{Stark et~al.}{2010}]{Stark10}
Stark, D.~P., Ellis, R.~S., Chiu, K., Ouchi, M., and Bunker, A.: 2010,
\newblock {\em {MNRAS}} {\bf 408}, 1628

\bibitem[\protect\astroncite{Stark et~al.}{2011}]{Stark11}
Stark, D.~P., Ellis, R.~S., and Ouchi, M.: 2011,
\newblock {\em {ApJL}} 728

\bibitem[\protect\astroncite{Stark et~al.}{2007}]{Stark07}
Stark, D.~P., Loeb, A., and Ellis, R.~S.: 2007,
\newblock {\em {ApJ}} 668

\bibitem[\protect\astroncite{{Tilvi} et~al.}{2010}]{Tilvi10}
{Tilvi}, V., {Rhoads}, J.~E., {Hibon}, P., et~al.: 2010,
\newblock {\em ApJ} {\bf 721}, 1853

\bibitem[\protect\astroncite{Vanzella et~al.}{2009}]{Vanzella09}
Vanzella, E., Giavalisco, M., Dickinson, M., et~al.: 2009,
\newblock {\em {ApJ}} {\bf 695}, 1163

\bibitem[\protect\astroncite{Vanzella et~al.}{2011}]{Vanzella11}
Vanzella, E., Pentericci, L., Fontana, A., et~al.: 2011,
\newblock {\em {ApJL}} 730

\bibitem[\protect\astroncite{{Verhamme} et~al.}{2012}]{Verhamme12}
{Verhamme}, A., {Dubois}, Y., {Blaizot}, J., et~al.: 2012,
\newblock {\em {A\&A}} {\bf 546}, A111

\bibitem[\protect\astroncite{{Yajima} et~al.}{2012}]{Yajima12}
{Yajima}, H., {Li}, Y., {Zhu}, Q., and {Abel}, T.: 2012,
\newblock {\em {MNRAS}} {\bf 424}, 884

\bibitem[\protect\astroncite{Zheng et~al.}{2010}]{Zheng:2009ax}
Zheng, Z., Cen, R., Trac, H., and Miralda-Escude, J.: 2010,
\newblock {\em ApJ} {\bf 716}, 574

\end{thebibliography}
	
\end{document}